\documentclass[10pt,epsf,letterpaper]{article}

\usepackage{amsmath}
\usepackage{amsfonts}
\usepackage{verbatim}
\usepackage{amssymb}
\usepackage{graphicx}
\usepackage{mathrsfs}
\usepackage{physics}
\usepackage{hyperref}
\usepackage[utf8]{inputenc}

\hypersetup{colorlinks=true,	linkcolor=blue,	filecolor=magenta, 	urlcolor=blue,	citecolor=blue,	pdftitle={Anomalous changing of geodesics in hairy black holes}, pdfpagemode=FullScreen}

\providecommand{\U}[1]{\protect\rule{.1in}{.1in}}

\newcommand{\br}{\biggr}

\graphicspath{{./fig_crop/}}
\renewcommand{\(}{\left(}

\renewcommand{\[}{\left[}

\usepackage{subcaption} %multple captions figures
\usepackage[title]{appendix}
\usepackage{multirow}
\usepackage[numbers, sort]{natbib} %natbib
\usepackage{titlesec}
\usepackage{etoolbox, chngcntr}
\AtBeginEnvironment{appendices}{%
 \titleformat{\section}{\bfseries\Large}{\appendixname~\thesection:}{0.5em}{}%
 \titleformat{\subsection}{\bfseries\large}{\thesubsection}{0.5em}{}%
\counterwithin{equation}{section}
}
\usepackage{floatrow}
\begin{document}

%%%%%%%%%%%%%%%%%%%%%%%%%%%%%%%%%%%%%%%%%%%%%%%%%%%%%%%%%%%%%%%%%%%%%%%%%%%%%%%%%%%%%%%%%%%%%%%%

\title{Anomalous changing of geodesics in hairy black holes}

\author{Weyner Ccuiro$^{(1)}$, David Choque$^{(2)}$ and Gustavo Valdivia-Mera$^{(3)}$\\
\\\textit{$^{(1)}$Universidad Nacional de San Antonio Abad del Cusco,} \\\textit{Av. La Cultura 733, Cusco, Per\'u.}
\\\\\textit{$^{(2)}$Pontificia Universidad Cat\'olica de Valpara\'\i so, Instituto de F\'\i sica,} \\\textit{Av. Brasil 2950, Valpara\'{\i}so, Chile.} 
\\\\\textit{$^{(3)}$East African Institute for Fundamental Research (ICTP-EAIFR),} \\\textit{University of Rwanda, Kigali, Rwanda.} }

\maketitle
\begin{abstract}
	We study the motion of test particles and the propagation of light around neutral hairy black holes under the influence of a self-interacting real scalar field minimally coupled to gravity. The goal of the present work is to show that the time-like and null-like geodesics have an anomalous behaviour for a special range of parameters in the dense hair region, defined as $\sqrt{\Omega(x_{h})}\leq r\leq 2MG_{N}/c^{2}$.

\end{abstract}
{\hypersetup{linkcolor=black}
\tableofcontents}

\newpage

\section{Introduction}\label{sec:Intro}

Hairy solutions are extensively constructed in the context of different theories, in some cases with minimal and non-minimal coupling with the Einstein-Hilbert theory \cite{Herdeiro:2014jaa,Anabalon:2017yhv,Astefanesei:2020qxk,Anabalon:2020pez,henneaux2002black,correa2012hairy,martinez2004exact,martinez2006electrically,martinez2003sitter,martinez2006topological}. In addition, there are many exact and numerically solutions for higher curvature theories \cite{Junior:2021atr,Santos:2021nbf,Herdeiro:2021lwl,Astefanesei:2020qxk,Astefanesei:2019qsg}. Another relevant topic is the thermodynamic of hairy black holes \cite{Anabalon:2015ija,Anabalon:2016izw,Astefanesei:2019ehu,Astefanesei:2019qsg,Anabalon:2019tcy,Astefanesei:2020xvn} where there are an interesting window of parameters in which the asymptotically flat hairy black holes are stable \cite{Astefanesei:2019mds}. In the present work, we focus on the minimal coupling with real scalar field. It was constructed in \cite{Acena:2013jya,Anabalon:2013qua,Anabalon:2013baa}, and consists of a general hairy family of asymptotically AdS solutions. Surprisingly if we fix, at the level of theory, $\Lambda=0$ we get an exact hairy black hole solution which is asymptotically flat. That hairy solutions was studied extensively in \cite{Anabalon:2013sra,Acena:2013jya}, and it can be embedded in SUGRA theories  \cite{Anabalon:2013eaa,Anabalon:2020qux,Anabalon:2020pez}. The hairy solutions presented here can evade the no-hair theorem \cite{Hertog:2006rr}, and the stability of the present hairy black holes is ensured by the scalar potential and its extreme points. Clearly, if I have only a kinetic term of the (real) scalar field in the theory, would not possible to evade the no-hair theorem, but if we add a non-minimal coupling gauge field it can behave like a scalar potential and there will be a possibility to evade the no-hair theorem \cite{Astefanesei:2007vh}. When we have complex scalar fields or another exotic fields, it is required another details \cite{Herdeiro:2016tmi,Vincent:2016sjq}. 
%In the literature doesn't exist exact solutions of rotational hairy black holes.
Our purpose in the present article is consider an exact solution of one scalar field coupled to a static black hole.\\

An important motivation is the study the posible anomalous geodesics in hairy black holes. This family of hairy black hole solutions has a horizon radius $r_{h}$ which is completely different from Schwarzschild radius $2MG_{N}/c^{2}$. We will show a concrete realization previously founded in \cite{Nunez:1996xv}, indeed we can prove that $r_{h}\leq 2MG_{N}/c^{2}$, then exist a region $\mathcal{D}\equiv 2MG_{N}/c^{2}-r_{h}$ called the dense-hair region. That region has an important effect on the geodesics due to the backreaction of scalar field, which causes a geodesic anomalous deflection. In the figures of the black hole horizon, we use grey color to show the Schwarzschild region and black, for the hairy black hole. So, the region between them is the dense hair region $\mathcal{D}$. This simple and novel property has an interesting effect on geodesic configurations, which is highly relevant because it allows us, in future works, to build models that can be tested by analysis of the black hole's shadows, e.g. the Event Horizon Telescope: \textit{The Shadow of the supermassive Black Hole} \cite{event2019first} and \textit{The Shadow and Mass of the Central Black Hole} \cite{akiyama2019event}.\\

The qualitative nature will be studied in \cite{david1511}, and here we construct numerically the time-like and null-like trajectories. We use the Runge-Kutta method of fourth-order(RK4), which means that there are four parameters to obtain. The numerical plots and calculations were made in Python 3.8 \cite{10.5555/1593511} with the following libraries, NumPy \cite{harris2020array}, SciPy \cite{2020SciPy-NMeth}, Matplotlib \cite{Hunter:2007}. All the plots of hairy black holes have been made using the library \textit{hairyBH} \cite{weyner:2021} and for Schwarzschild see \cite{gustavo:2021}.\\

The present work is organized in the following form: First, in section \ref{sec:2}, we consider a brief description of geodesics of Schwarzschild black hole, and its respective details are in the appendix \ref{ap:1}. 
In section \ref{sec:3}, we describe the theory and properties of the hairy solution, such as the horizon existence and the mass for each branch. In section \ref{sec:4} and \ref{sec:5}, we construct the orbital equation and solve it numerically. In addition, we present the plots of the effective potential and its respective trajectories for each region. In section \ref{sec:6}, we construct the near horizon geometry for hairy black holes and we solve the geodesics equations, we verify the interesting anomalous changing of the geodesics shown in the following figures: \ref{fig7:2} and \ref{fig11a}. Finally, in section \ref{sec:7}, we present the discussion and future directions.

%%%%%%%%%%%%%%%%%%%%%%%%%%%%%%%%%%%%
\section{Geodesics of Schwarzschild black hole}
\label{sec:2}
\begin{figure}
	\begin{subfigure}{.5\textwidth}
	  \centering
	  \includegraphics[width=.8\linewidth]{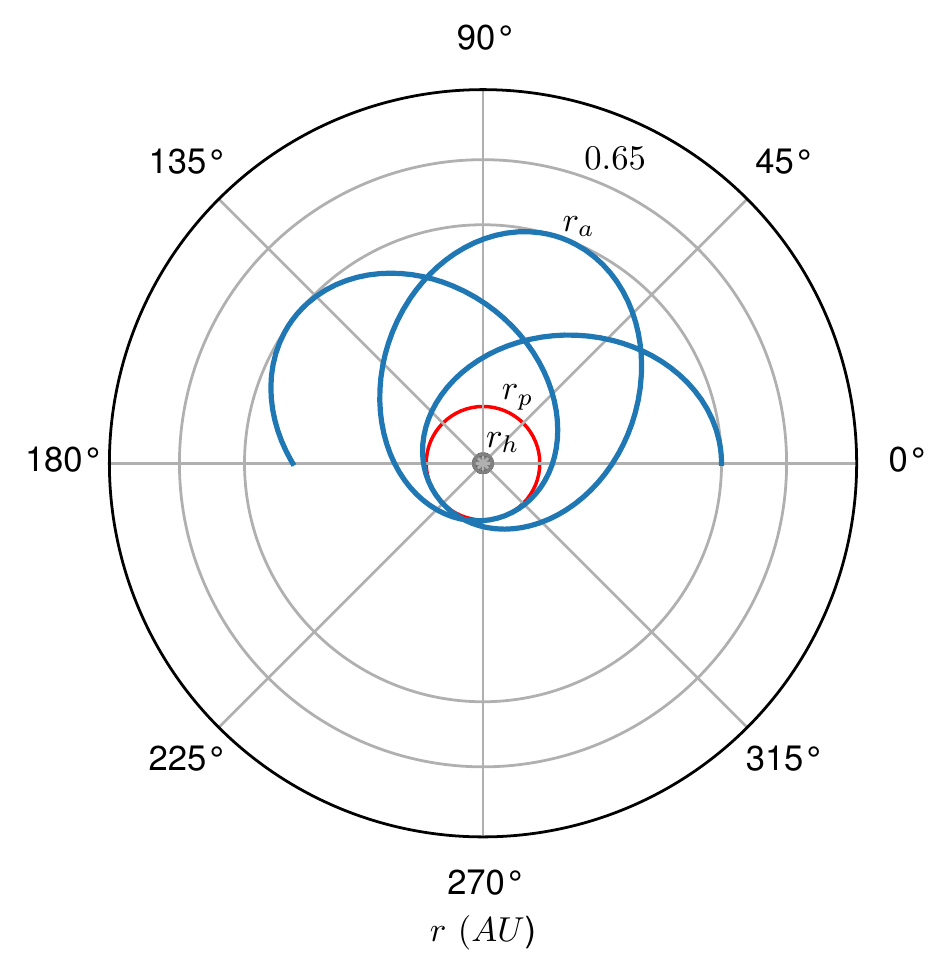}
	  \caption{}
	  \label{Schwa1:fig1}
	\end{subfigure}%
	\begin{subfigure}{.5\textwidth}
	  \centering
	  \includegraphics[width=.8\linewidth]{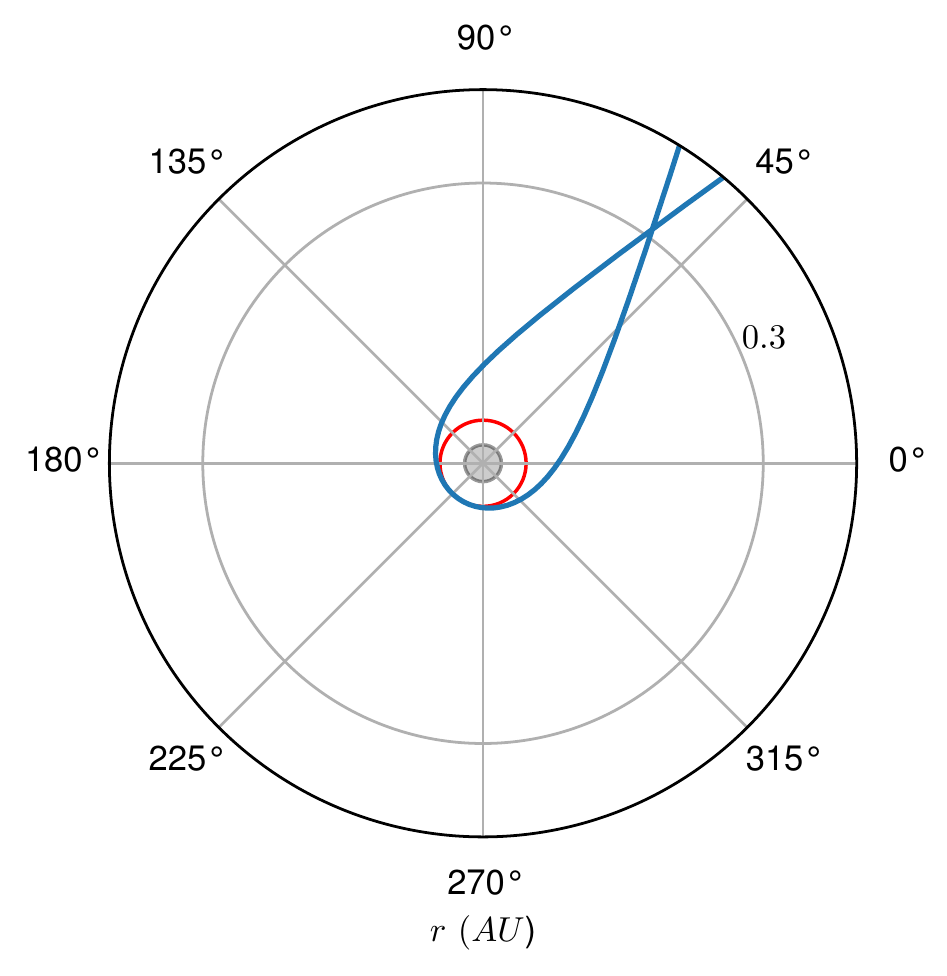}
	  \caption{}
	  \label{Schwa1:fig2}
	\end{subfigure}
	\caption{\small\textbf{ (Time-like)}\\
	Here we consider: $r_h$ (grey), $r_p$ (red) and the orbit (blue). 
    Where $r_p$ indicates the radius of the minimum distance from the center of the black hole to the orbit.\\
	(\subref{Schwa1:fig1}): For energy $E=E_2=-0.03$ and initial conditions: $r_0=0.51069\,\,AU$ and $\dot{r}_0=0$. Where, $r_0=r_a$ indicates the apoapsis, while $r_p=0.12158\,\,AU$, the periapsis. According to (\ref{devi1}) the angular deviation of the apoapsis per orbital period is $\Delta\varphi=61.1^{0}$, you can verify that in this figure.\\
	(\subref{Schwa1:fig2}): $E=E_3=0.2$, initial conditions: $r_0\rightarrow\infty$ and $\dot{r}_0=\sqrt{Ec^2}$. Where, $r_2=0.04613\,\,AU$ indicates the radius of the minimum distance from the center of the black hole to the orbit.}
	\label{Schwa1}
\end{figure}

\begin{figure}
	\begin{subfigure}{.5\textwidth}
	  \centering
	  \includegraphics[width=.8\linewidth]{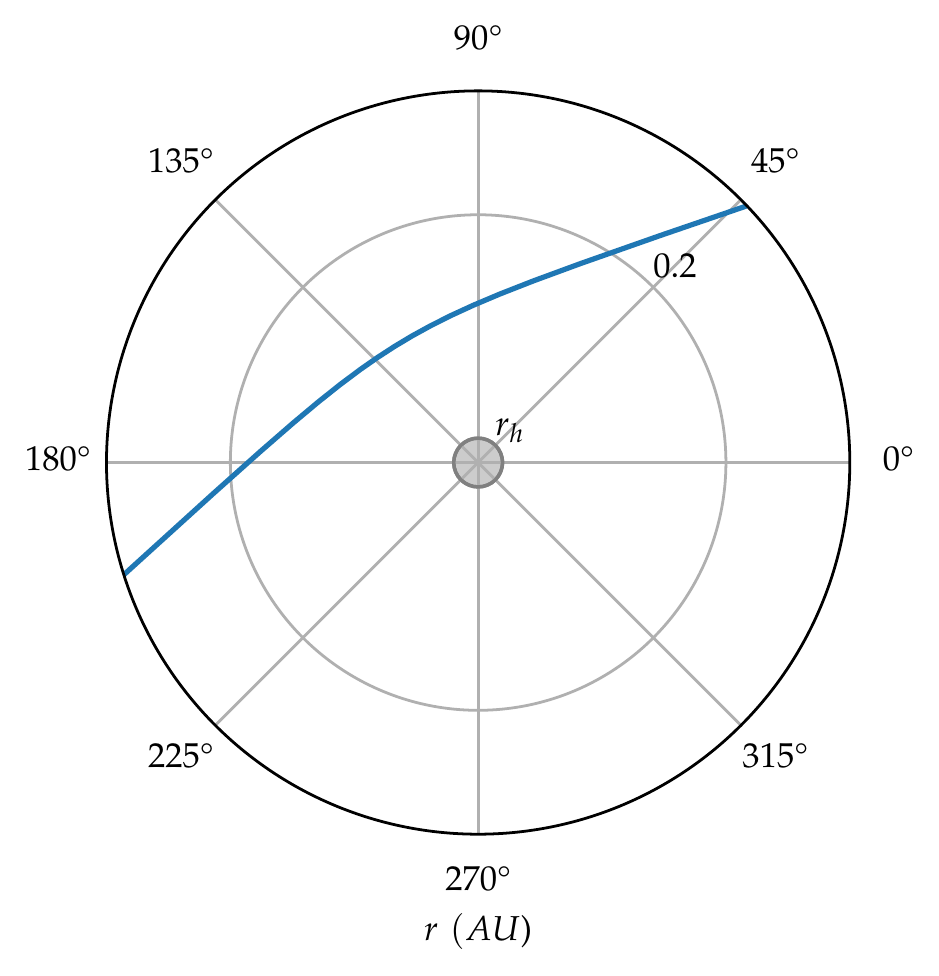}
	  \caption{$r_h$ and the orbit (blue).}
	  \label{timee:fig1}
	\end{subfigure}%
	\begin{subfigure}{.5\textwidth}
	  \centering
	  \includegraphics[width=.8\linewidth]{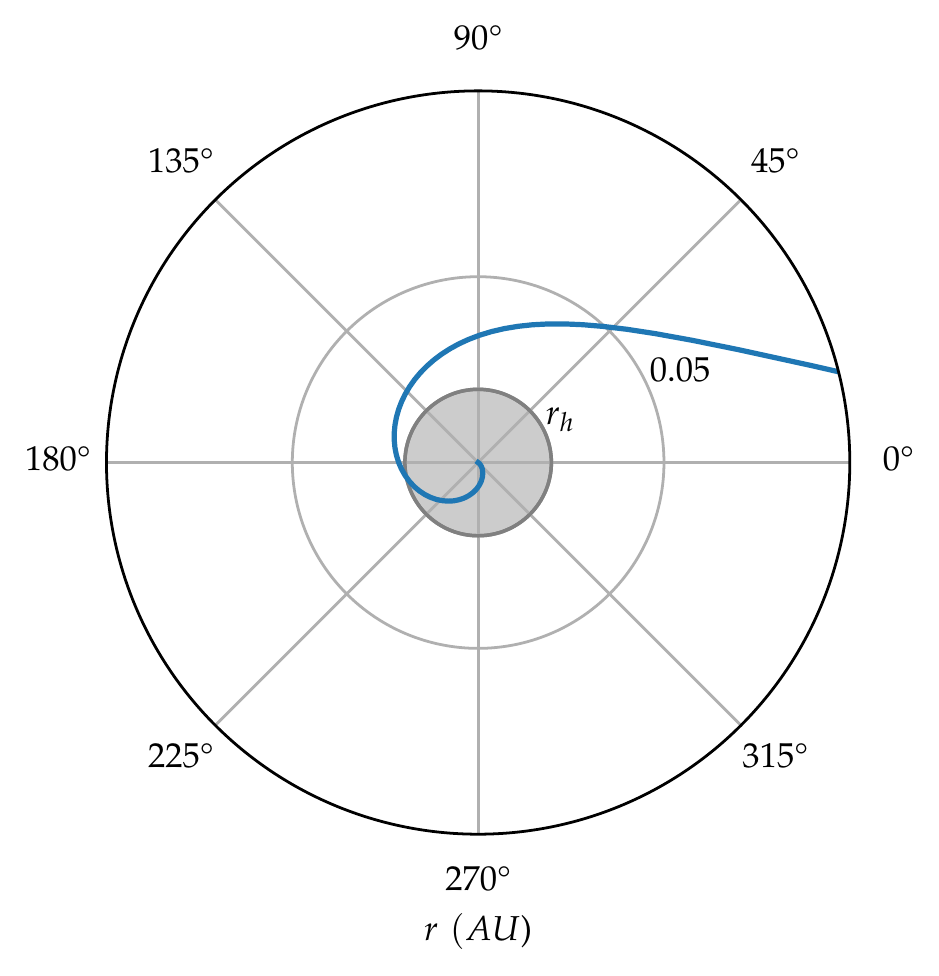}
	  \caption{$r_h$ and the orbit (blue).}
	  \label{timee:fig2}
	\end{subfigure}
	\caption{\small\textbf{ (Null-like)}\\(\subref{timee:fig1}): $E=E_1=10^{21} AU^2 yr^{-4}$, initial conditions: $r_0\rightarrow\infty$ and $\left(\frac{dr}{d\xi}\right)_0=\sqrt{E}$. The impact parameter is given by $b^2=3.9993\times10^{-12}\,\,yr^2$. The light deviation can be calculated from (\ref{desvi1}), we have $R_{0}=0.1188965650~AU$, them $\vert\Delta\varphi-\pi\vert\frac{180^{0}}{\pi}\approx 22^{0}$.\\
    (\subref{timee:fig2}): $E=E_1=7.5\times 10^{21} AU^2 yr^{-4}$, initial conditions: $r_0\rightarrow\infty$ and $\left(\frac{dr}{d\xi}\right)_0=\sqrt{E}$. The impact parameter is given by $b^2=5.3324\times10^{-13}\,\,yr^2$.}
	\label{timee}
\end{figure}

%%%%%%%%%%%%%%%%%%%%%%%%%%%%%%%%%
%
All the details of the present section can be found in appendix \ref{apendice1}. The purpose of the present section is to show the geodesics of the Schwarzschild black hole and compare it with the hairy case. % , in special we compare the horizon radius which for Schwarzschild (grey region) and the hairy solution are respectively $r_{h}=\frac{2G_{N}M}{c}$ and $\sqrt{\Omega}$.
It is interesting that the Schwarzschild horizon radius is bigger than the radius of the hairy solution, and it means that the scalar field (hair) allows the existence of more compact objects. The geodesics for hairy black holes are studied in section \ref{sec:2}.
First, we show the null (see Figure \ref{Schwa1}) and time-like (see Figure \ref{timee}) geodesics for Schwarzschild, where we have used $M=10^6 M_{\odot}$, $\bar{\mathcal{J}}^2=5.82708\times 10^{-13} yr^2$ and $r_h\approx 0.019708\,\,AU$.
The numerical construction of geodesics is extensively 
know in the literature \cite{Junior:2021atr,Santos:2021nbf,Herdeiro:2021lwl,Anabalon:2020loe}.

%r_0=0.5106936776245934
%r_p=0.12158224399382238
%r_2=0.04613484978865398
%%%%%%%%%%%
%%%%%%%%%%%

In figure (\ref{timee:fig2}) the geodesics fall into the black hole (Schwarzschild case), across the horizon, in a tangential form.

%%%%%%%%%%%%%%%%%%%%%%%%%%%%%%%%%%%%
\section{Hairy Black hole solution}
\label{sec:3}
%%%%%%%%%%%%%%%%%%%%%%%%%%%%%%%%%%%%%
%
We consider the following modified Einstein-Hilbert theory with an scalar field with non-minimal coupling %
\begin{equation}
  I[\textbf{g},\phi]=\frac{1}{2\kappa}\int_{\mathcal{M}}d^{4}x\sqrt{-g}{R}+\int_{\mathcal{M}}d^{4}x\sqrt{-g}\left[-\frac{1}{2}(\partial\phi)^{2}-V(\phi)\right]
  \label{action}
\end{equation}
here the coupling constant is $\kappa=\frac{8\pi G_{N}}{c^{4}}$, where $G_N$ is the constant of gravitation and $c$ is the speed of light, $g$ is the determinant of the metric tensor, $g=\det (g_{\mu\nu})$, $R$ is the Ricci scalar and $V(\phi)$ is the self-interaction term.
The equations of motion arising from the action principle are
\begin{equation}
  G_{\mu\nu}
  =\kappa\,T_{\mu\nu}, \quad
  \frac{1}{\sqrt{-g}}\,\partial_{\mu}
  \left(\sqrt{-g}g^{\mu\nu}\partial_{\nu}\phi\right)
  =\frac{dV}{d\phi}
\end{equation}
where the Einstein tensor and the energy-momentum tensor for the scalar field are respectively 
\begin{equation}
  \begin{split}
    G_{\mu\nu} & :=R_{\mu\nu}-\frac{1}{2}g_{\mu\nu}R \\
    T_{\mu\nu} & =
    \partial_{\mu}\phi\,\partial_{\nu}\phi
    -g_{\mu\nu}\left[\frac{1}{2}\left(\partial\phi\right)^2
      +V(\phi)\right]
  \end{split}
\end{equation}
Following to \cite{Anabalon:2013eaa,Anabalon:2017yhv,Anabalon:2012ih,Anabalon:2016izw}, 
we consider the exotic potential $V(\phi)$, it presents a non-trivially self-interaction, it was first obtained and presented in \cite{Anabalon:2013eaa},
\begin{equation}
  V(\phi)=\frac{\alpha}{\kappa\nu^2}\biggl{\{}\frac{\nu-1}{\nu+2}\sinh{\left[l_\nu(\nu+1)\phi\right]}-\frac{\nu+1}{\nu-2}\sinh{[l_{\nu}(\nu-1)\phi]}+4\left(\frac{\nu^{2}-1}{\nu^{2}-4}\right)\sinh{(l_\nu\phi)}\biggr{\}}
  \label{pot1}
\end{equation}
where
$l_{\nu}\equiv\left(\frac{2\kappa}{\nu^{2}-1}\right)^{1/2}$. This theory has two novels parameters, $\alpha$, which has an important role in the existence of the horizon and $\nu$, which can calibrate the scalar field $\phi$.
Considering the following ansatz for conformal metric
\begin{equation}
  ds^{2}=\Omega(x)\left[  -c^{2}f(x)dt^{2}+\frac{\eta^{2}dx^{2}}{f(x)}+d\theta
  ^{2}+\sin^{2}\theta d\varphi^{2}\right]  \label{Ansatz}%
\end{equation}
we can integrate the equations of motion for the metric and scalar field, such that we  obtain the family of hairy solutions \cite{Anabalon:2013sra,Anabalon:2013qua,Acena:2012mr,Acena:2013jya},
\begin{equation}
  \phi(x)=l_{\nu}^{-1}\ln{x}
  \label{campo}
\end{equation}
where the conformal factor $\Omega(x)$ and the metric function $f(x)$ are given by
\begin{equation}
  \begin{split}
    \Omega(x) & =\frac{\nu^{2}x^{\nu-1}}{\eta^{2}(x^{\nu}-1)^{2}} \\
    f(x) & =\alpha\biggl{[}\frac{1}{\nu^{2}-4}-\frac{x^{2}}{\nu^{2}%
    }\biggl{(}1+\frac{x^{-\nu}}{\nu-2}-\frac{x^\nu}{\nu+2}%
    \biggr{)}\biggr{]}+\frac{x}{\Omega(x)}
    \label{omega}
  \end{split}
\end{equation}
Actually, there are two branches of spacetime in which the physical quantities are well defined. The quantity $\eta$ is a positive definite constant of integration that is related to the mass of the black hole\footnote{Along the paper, we mostly use the unit system where the constants have the following values:
  \begin{equation}
    \begin{split}
      G_N & \approx39.409\,{AU}^{3}{M}^{-1}_{\odot}{yr}^{-2}, \\
      c & \approx 6.324\cdot 10^{4}\,{AU}\,{yr}^{-1},\\
      \kappa & \approx 6.192\cdot 10^{-17}
      \,{AU}^{-1}{M}^{-1}_{\odot}yr^2
    \end{split}
  \end{equation}
  where ${M}_\odot$ represents the solar mass, $AU$ is for the Astronomical Unit and $yr$ stands for a year}. The dimension of the parameter and constant of integration are $\dim\alpha=\dim\eta^{2}=\text{length}^{-2}$ and $\dim\eta=\text{length}^{-1}$. In addition, $\nu$ is a dimensionless parameter.\newpage The principal characteristic of each branch are
\begin{itemize}
  \item {\bf{Negative Branch}}:
        The coordinate of the black hole horizon $x_{h}$ is less than 1, then the range of the coordinate $x$ is given by $x_{h}<x<1$. The boundary is located at $x=1$ and the singularity at $x=0$. In this case the scalar field is negative definite $\phi<0$
  \item {\bf{Positive Branch}}: The coordinate of the black hole horizon $x_{h}$ is greater than 1, then the range of the coordinate $x$ is given by  $1<x<x_{h}$. The boundary is located at $x=1$ and the singularity at $x=\infty$. Here the scalar field becomes positive definite $\phi>0$
\end{itemize}
The radial coordinate system is more
intuitive than conformal metric (\ref{Ansatz}).
The equation $r^2=\Omega(x)$, which relates both coordinate systems, cannot be solved exactly but is easy to get the asymptotic  coordinate transformation \cite{Anabalon:2015xvl}
\begin{equation}
  \label{Omeg}
  x= 1\pm \frac{1}{\eta r}
  \mp\frac{\nu^2-1}{24\eta^3r^3}\left[1
    \mp\frac{1}{\eta r}
    \mp\frac{9(\nu^2-9)}{80\eta^2r^2}\right]
  +\mathcal{O}(r^{-6})
\end{equation}
In \cite{Astefanesei:2005ad,Astefanesei:2009wi,Astefanesei:2006zd,Astefanesei:2010bm,Anabalon:2015xvl,Anabalon:2014fla} you can find a quasilocal formalism used to find the mass (energy) of the gravitational system, and for asymptotically AdS spacetime in the presence of the scalar field \cite{Astefanesei:2009wi}.
The scalar field is a secondary hair, therefore, there is not a constant of integration associated to it, so, we have a unique constant of integration $\eta$ which is just related with the mass. The ADM mass can be read-off from $g_{tt}$ in canonical coordinates, in \cite{david1511} we will construct 
a quasilocal stress tensor\footnote{For asymptotically AdS space-time here you can find an interesting discussions \cite{Skenderis:2002wp,deHaro:2000vlm}} (asymptotically flat) in order to get $M$, see \cite{Astefanesei:2019mds,Kraus:1999di}
\begin{equation}
  -\frac{1}{c^2}g_{tt}=
  1-\frac{2G_{N}M}{c^2\,r}+\mathcal{O}(r^{-3})
\end{equation}
where the masses for negative branch and positive branch are respectively 
\begin{equation}
  M=\frac{c^2}{2G_{N}}
  \left(\frac{\alpha+3\eta^2}{3\eta^3}\right), \qquad \phi\leq 0 
  \label{mass1}
\end{equation}
\begin{equation}
  M=-\frac{c^2}{2G_{N}}
  \left(\frac{\alpha+3\eta^2}{3\eta^3}\right),
  \qquad \phi\geq 0
  \label{mass2}
\end{equation}
\subsection{Evading the no hair theorem}
\begin{figure}
	\begin{subfigure}{.5\textwidth}
	  \centering
	  \includegraphics[width=.8\linewidth]{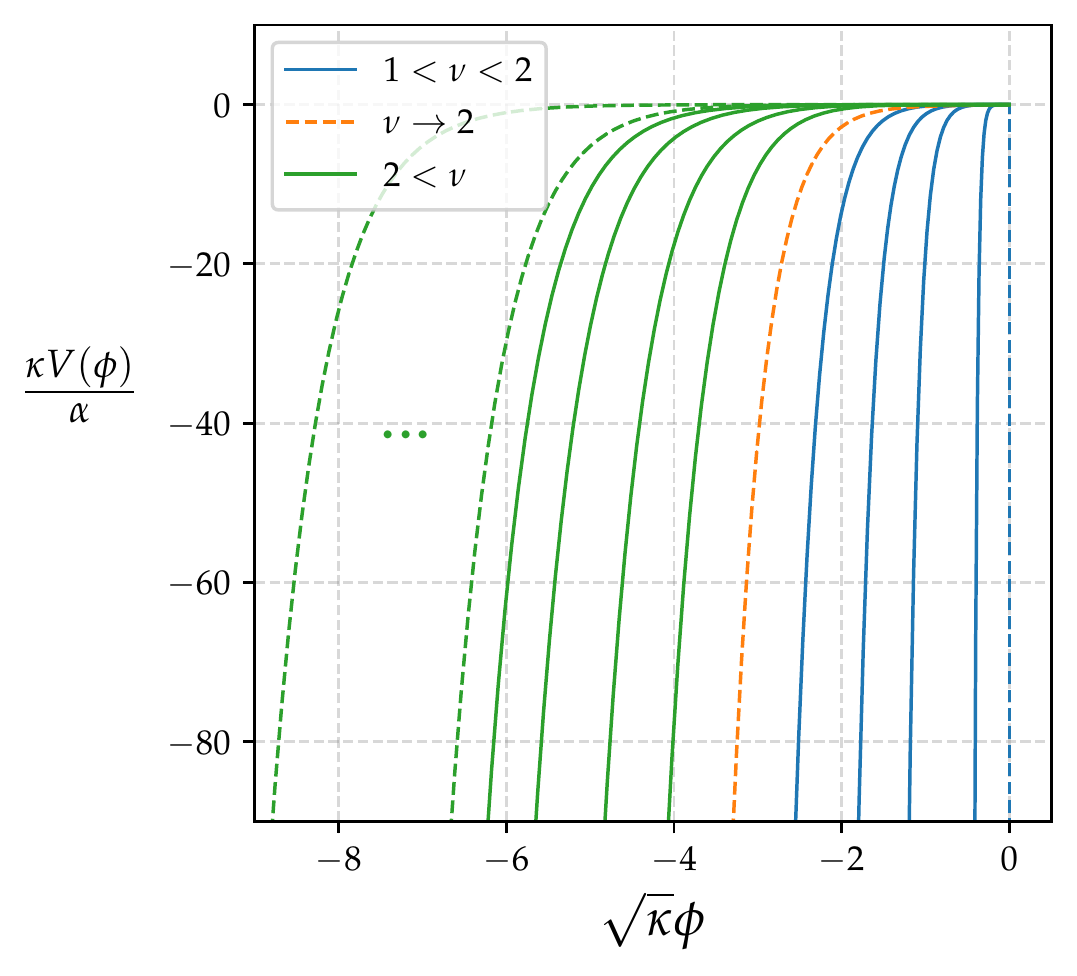}
	  \caption{}
	  \label{gttneg-A}
	\end{subfigure}%
	\begin{subfigure}{.5\textwidth}
	  \centering
	  \includegraphics[width=.8\linewidth]{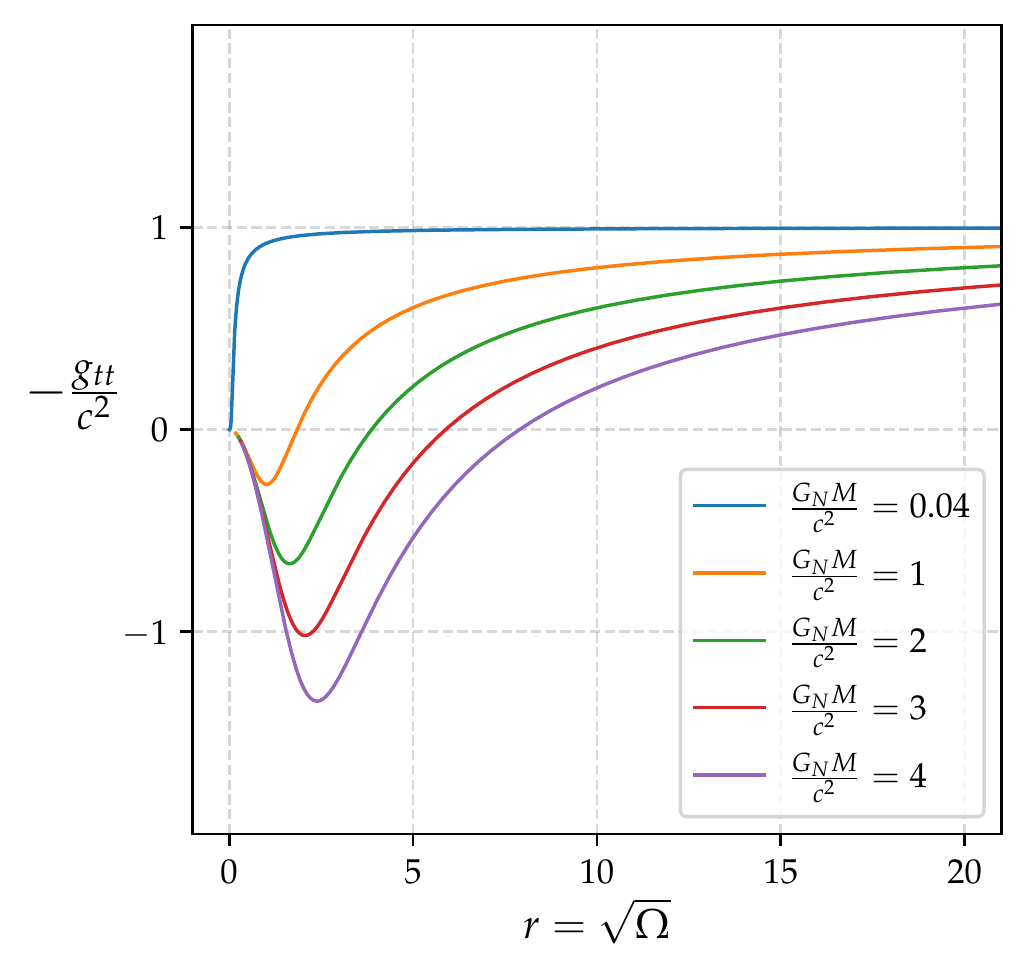}
	  \caption{}
	  \label{gttneg-B}
	\end{subfigure}
	\caption{\small Negative branch. (\subref{gttneg-A}):
    $\frac{\kappa}{\alpha}V(\phi)~vs~\sqrt{\kappa}\phi$. (\subref{gttneg-B}): $-g_{tt}/c^{2}~vs~\sqrt{\Omega}$: We consider the hairy parameter $\nu=1.52$, $\alpha=1~AU^{-2}$ and the range of masses for the black holes $0.04~AU\leq\frac{GM_{N}}{c^{2}}\leq 4.00~AU$.}
	\label{fig:neg}
\end{figure}

The no hair theorem can be evaded if we have a potential with a global maximum at the boundary and a minimum at the horizon. That condition is ensured by $d^2V/d\phi^2\leq 0$. Considering the following quantities for the spacetime $0\leq x\leq 1$ which is named as negative branch $\phi\leq 0$:
\begin{itemize}
  \item The potential $V(\phi)$ depends on the parameters $\nu$ and $\alpha$. From figure (\ref{fig:neg}), we check that $d^2V/d\phi^2\leq 0 ~\Leftrightarrow~ \nu>1,~~\alpha>0$, and the scalar potential has a global minimum at the horizon $V(\phi_{h})$.
  \item From figure (\ref{fig:neg}), the horizon existence, $-\frac{g_{tt}}{c^2}=0$, is ensured if $\alpha>0$, and there is an additional condition for the mass:
        \begin{itemize}
          \item For: $1\leq\nu<2$ there is no restriction on the mass of the black hole.
          \item For: $2<\nu$ in \cite{Anabalon:2017yhv} they showed an interesting lower bound
                for the mass of hairy black holes which are asymptotically AdS, and here we present our result for asymptotically flat case. Clearly if the scalar field increase (that is dominated by hairy parameter $\nu$) the black hole mass has a minimum value which can hold up the horizon
                \begin{equation}
                  \label{macri1}
                  M>M_{cri}\equiv\frac{c^{2}}{2G_{N}}\qty(\frac{\nu-2}{\alpha})^{1/2}
                \end{equation}
        \end{itemize}
\end{itemize}
\begin{figure}
	\begin{subfigure}{.5\textwidth}
	  \centering
	  \includegraphics[width=.8\linewidth]{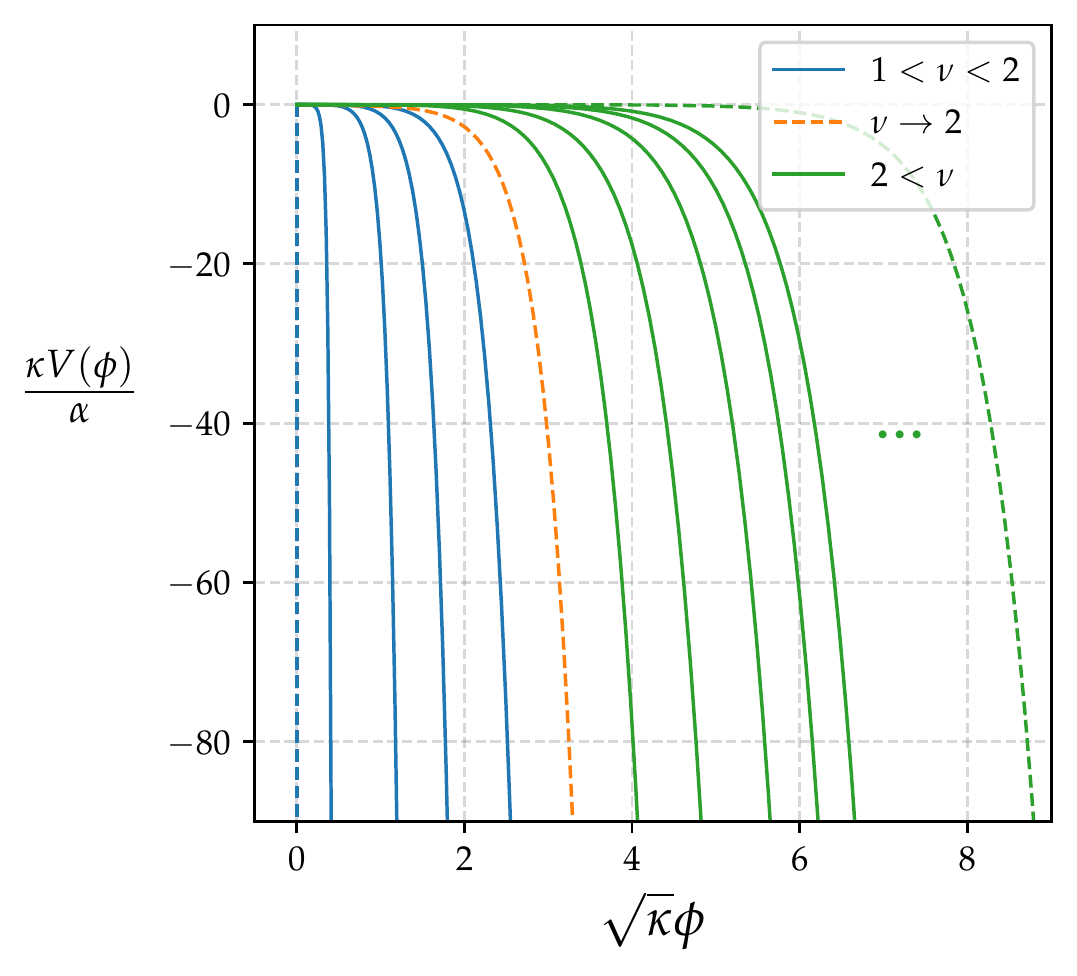}
	  \caption{}
	  \label{gttpo}
	\end{subfigure}%
	\begin{subfigure}{.5\textwidth}
	  \centering
	  \includegraphics[width=.8\linewidth]{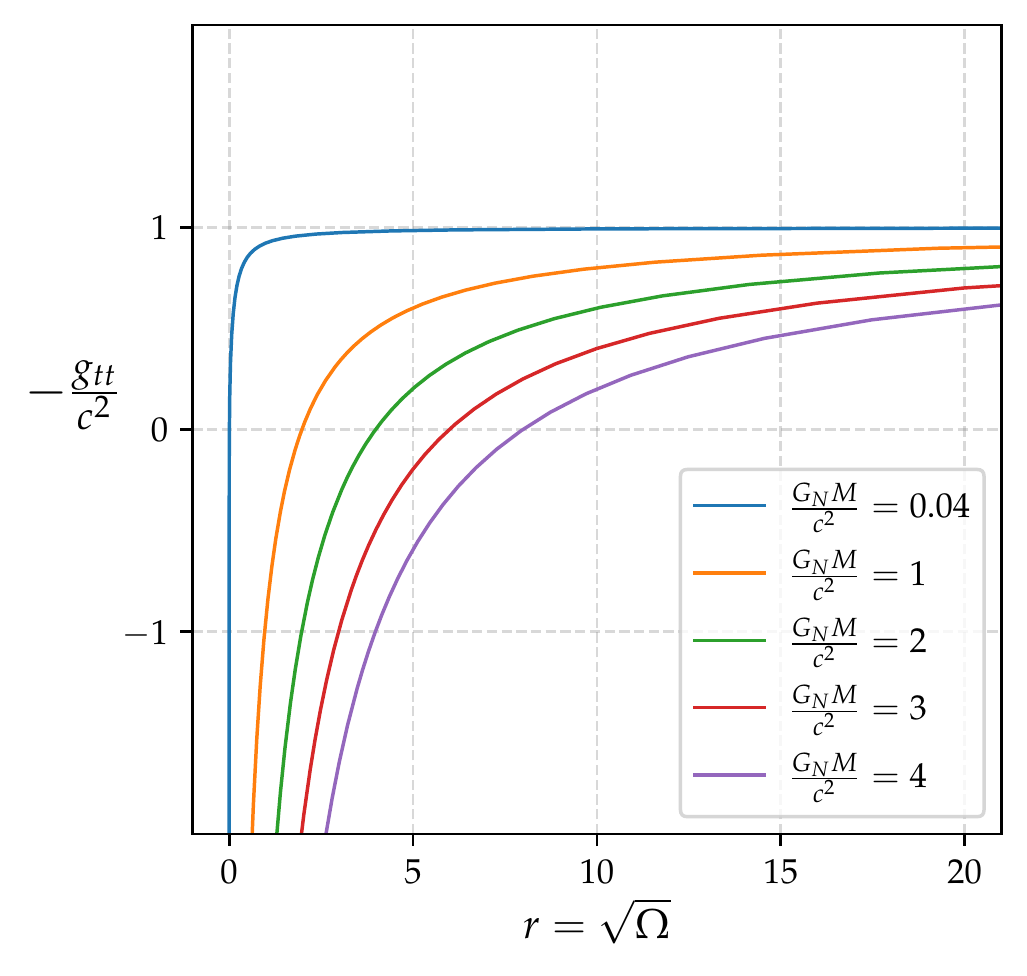}
	  \caption{}
	  \label{fig:po}
	\end{subfigure}
	\caption{\small Positive branch. (\subref{gttpo}):
    $\frac{\kappa}{\alpha}V(\phi)~vs~\sqrt{\kappa}\phi$. (\subref{fig:po}): $-g_{tt}/c^{2}~vs~\sqrt{\Omega}$, We consider the hairy parameter $\nu=1.76$, $\alpha=-40~AU^{-2}$ and the range of masses for the black holes $0.04~AU\leq\frac{G_{N}M}{c^{2}}\leq 4.00~AU$}
	\label{fig:fig}
\end{figure}

and for the spacetime $1\leq x\leq \infty$ we have the positive branch $\phi\geq 0$:
\begin{itemize}
  \item The potential $V(\phi)$ depends on the parameters $\nu$ and $\alpha$. From figure (\ref{fig:po}), we check that $d^2V/d\phi^2\leq 0 ~\Leftrightarrow~ \nu>1,~~\alpha<0$ and the scalar potential has a global minimum at the horizon $V(\phi_{h})$.
  \item From figure (\ref{fig:po}), the horizon existence, $-\frac{g_{tt}}{c^2}=0$, is ensured if $\alpha<0$, and for all $\nu>1$ exist a critical minimum mass
        \begin{equation}
          \label{macri2}
          M>M_{cri}\equiv\frac{c^{2}(\nu-1)(\nu+2)}{6G_{N}\sqrt{-\alpha(\nu+2)}}
        \end{equation}
\end{itemize}
The existence of minimal mas giving in (\ref{macri1}) for the negative branch ($\phi<0$), with $\nu>2$, can be interpreted in a similar form to Kerr-Black holes, in which the horizon existence is ensured by the inequality between the angular momentum density and mass of the black hole. In the hairy case, the horizon is ensured by the critical mass which is a function of the hairy parameters $\nu$ and $\alpha$. So, if the mass $M$ is not enough, the scalar field implodes and the horizon disappears.
The positive branch has the same interpretation, but in that case the minimal mass condition is given for the entire range of values of parameter $\nu$ $(1\leq\nu\leq\infty)$, which describes the back reaction of the scalar field.

The no hair limit can be obtained if the hair parameter
is fixed to $\nu = 1$. In consequence, the scalar field \eqref{pot1} is null and the metric \eqref{Ansatz} takes the following form

\begin{align}
  \Omega (x) = & \dfrac{1}{\eta^{2}\qty(x-1)^{2}}, \\
  f(x) = & \dfrac{1}{3}\alpha(x-1)^{3}+\eta^{2}x(x-1)^{2}
\end{align}
So, we can recover the Schwarzschild black hole in the canonical form by the following change of coordinates: Fixing $\nu=1$ in \eqref{Omeg}

\begin{equation}
  x = 1 \pm \dfrac{1}{\eta r}
  ~~\Rightarrow~~ 
 -g_{tt}= \Omega(x)f(x)\vert_{x(r)}=1-\frac{2G_{N}M}{c^{2}r}
\end{equation}

%%%%%%%%%%%%%%%%%%%%%%%%%%%%%%%%%%%%%%%%%%%%%%%%%%%%%%%%%%%%%
\section{Time-like geodesics}
\label{sec:4}
\begin{figure}
	\begin{subfigure}{.5\textwidth}
	  \centering
	  \includegraphics[width=.8\linewidth]{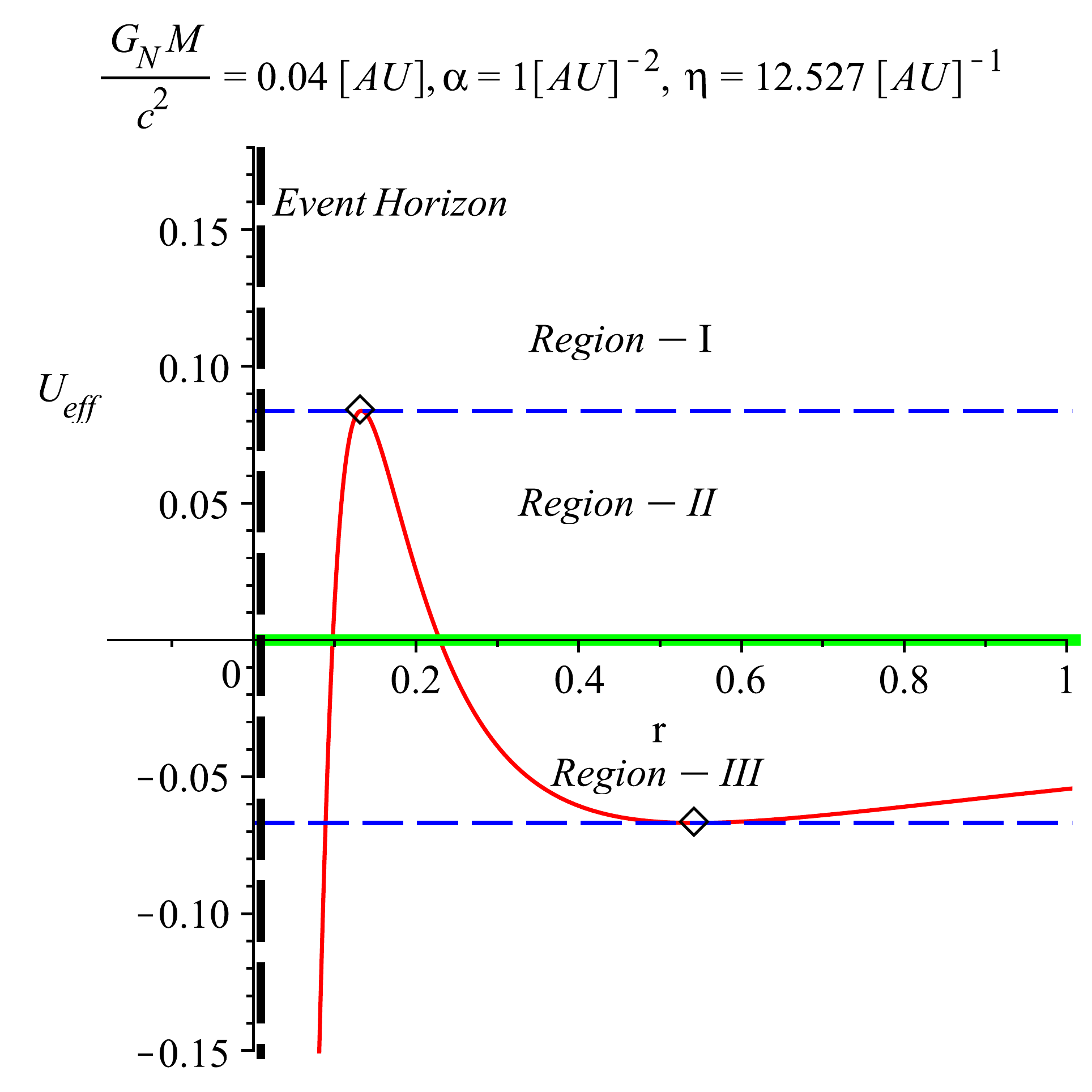}
	  \caption{Negative branch}
	  \label{fig5a}
	\end{subfigure}%
	\begin{subfigure}{.5\textwidth}
	  \centering
	  \includegraphics[width=.8\linewidth]{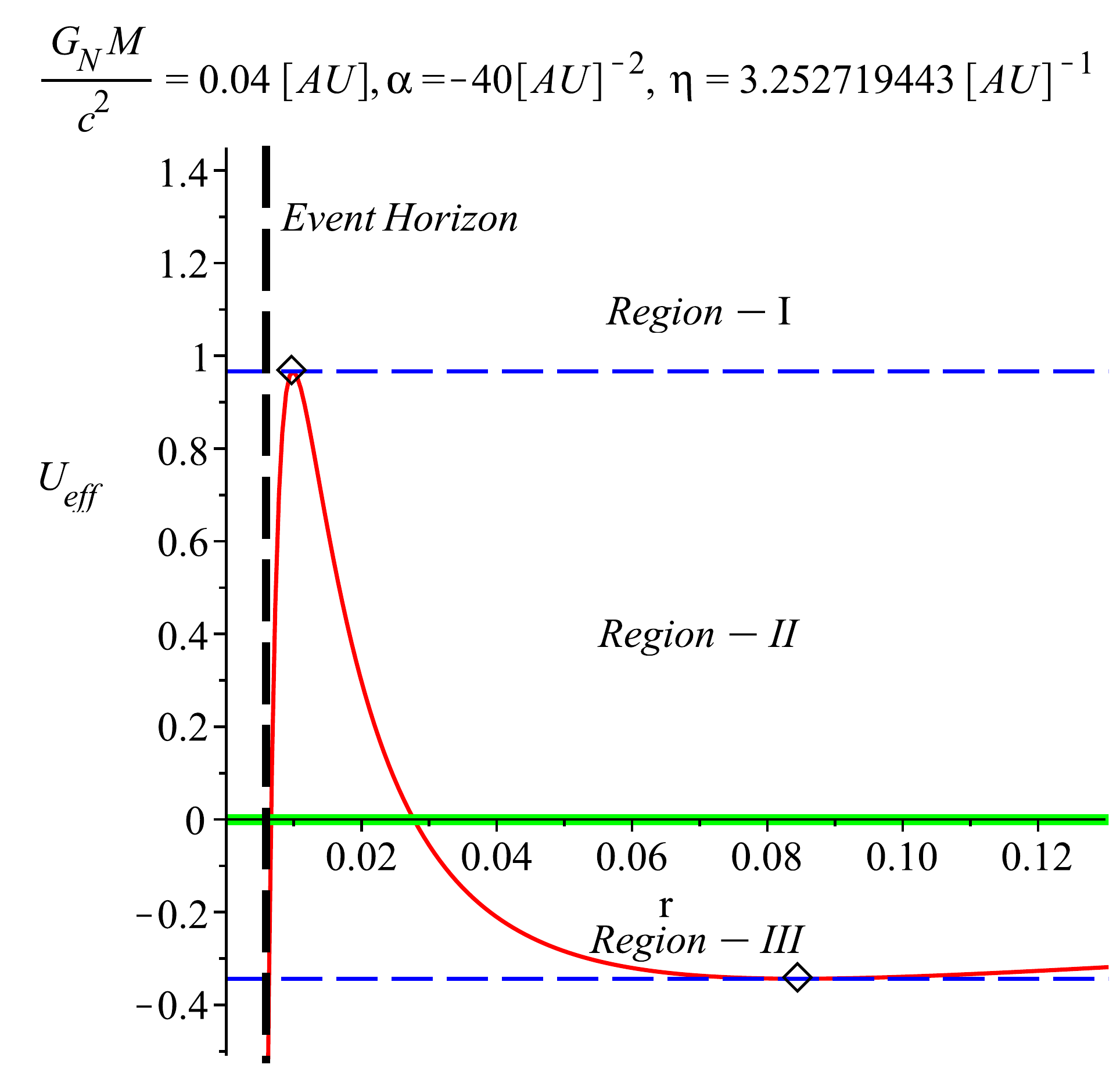}
	  \caption{Positive branch}
	  \label{fig5b}
	\end{subfigure}
	\caption{\small We have the positive-branch (\subref{fig5b}) and negative-branch (\subref{fig5a}) of the effective potential $U_{eff}$. The parameters are described in table (\ref{tab1}).}
	\label{Potential}
\end{figure}

%%%%%%%%%%%%%%%%%%%%%%%%%%%%%%%%%%%%%%%%%%%%%%%%%%%%%%%%%%%%%%%
%
In \cite{david1511} they will show the following
equations for the time-like orbits on the equatorial plane $\theta=\pi/2$, for hairy black hole solutions described in section \ref{sec:2}. The first order orbital equation is given by\footnote{Here the Killing vectors are
  \begin{equation}
    \frac{dt}{d\xi}=
    \frac{\bar{\mathcal{E}}}{\Omega(x)f(x)},~~ \frac{d\varphi}{d\xi}=\frac{\bar{\mathcal{J}}c^{2}}{\Omega(x)},  ~~
    \bar{\mathcal{E}}=\mathcal{E}/mc^2, ~~ \bar{\mathcal{J}}=\mathcal{J}/mc^{2}
  \end{equation}}
\begin{equation}
  \begin{split}
    \bar{\mathcal{E}}^{2}-1
    & =\left(\frac{\eta\Omega(x)}{c}\right)^2
    \dot{x}^{2}+U_{\text{eff}}(x), \\
    U_{\text{eff}}(x) & =\Omega(x)f(x)\left(1+\frac{\bar{\mathcal{J}^{2}}c^{2}}{\Omega(x)}\right)-1
    \label{pothairy1}
  \end{split}
\end{equation}
In order to get the second order equation 
we consider the following relations
\begin{equation}
  r=\sqrt{\Omega(x)}=r(x), \quad r(\varphi)=r(x) ~\Rightarrow~
  x(\varphi),~~ \varphi(\tau)
  \label{eq}
\end{equation}
%
%\clearpage
%
Taking the derivative of (\ref{pothairy1}) with respect to $\varphi$ we get a second order orbital equation which can be easily 
numerically solved\footnote{It is easy to show that with the following changes we can get the orbital equation for Schwarzschild $\nu=1$, see (\ref{geosh})
  \begin{equation}
    x(\varphi)=1-\frac{1}{\eta r(\varphi)}, \quad \alpha=3\eta^{3}r_{h}-3\eta^2, \quad r(\varphi)=\frac{1}{u(\varphi)}
  \end{equation}}
\begin{equation}
  \begin{split}
    \frac{d^{2}x}{d\varphi^{2}}+H(x,\bar{\mathcal{J}},\eta) & =0, \\
    H(x,\bar{\mathcal{J}},\eta) & =\frac{1}{2(\eta\bar{\mathcal{J}}c)^{2}}[(\Omega f)^{\prime}+(\bar{\mathcal{J}}c)^{2}f^{\prime}]
    \label{TMgeo}
  \end{split}
\end{equation}
remembering that the radial coordinate
is related to $x$-coordinate in the following exact form $r(x)=\sqrt{\frac{\nu^{2}x^{\nu-1}}{\eta^{2}(x^{\nu}-1)^{2}}}$, we can plot in polar coordinates $r(x)~\text{vs}~\varphi(x)$. The effective potential describes the following regions for negative and positive branch: Region-I:~ $\bar{\mathcal{E}}^{2}-1>U(r_{max})$. Region-II:~ $U(r_{max})>\bar{\mathcal{E}}^{2}-1>0$ and
Region-III:~$0>\bar{\mathcal{E}}^{2}-1>U_{min}$, see the Figure \ref{Potential}, a, b and see table (\ref{tab1}) for the hairy parameters. Where we consider $\bar{\mathcal{E}}=\mathcal{E}/mc^2$, and $\bar{\mathcal{J}}=\mathcal{J}/mc^{2}$. Also in Figure \ref{Potential} we can see the extremes of the effective potential $U_{eff}$. In this way we can set the parameters to plot a geodesic.
\begin{table}[ht]
  \centering
  \begin{tabular}{c c}
    \hline \hline
    \multirow{4}{1.3cm}{Negative Branch} & $x<1$,~$\alpha=1~AU^{-2}$,~ $\nu=1.52$                                                        \\ \cline{2-2}
                                         & $\eta\approx12.52655373~AU^{-1}$,~ $G_{N}M/c^{2}=0.04\,A|U$                                   \\\cline{2-2} &\\[-0.32cm] 
                                         & $\bar{\mathcal{J}}=2.6\times10^{-6} yr$,~$\bar{\mathcal{J}}_{c}\approx2.1072\times 10^{-6}yr$ \\ \cline{2-2}
                                         & $U_{max}\approx0.083747687$,~ $U_{min}\approx-0.0668590532$                                   \\ \hline
    \multirow{4}{1.3cm}{Positive Branch} & $x>1$,~ $\alpha=-40~AU^{-2}$,~ $\nu=1.76$                                                     \\\cline{2-2}
                                         & $\eta\approx3.252719443~AU^{-1}$,~ $G_{N}M/c^{2}=0.04\,AU$                                    \\\cline{2-2}
                                         & $\bar{\mathcal{J}}=7\times10^{-7} yr$,~$\bar{\mathcal{J}}_{c}\approx2.5280\times 10^{-7}yr$   \\\cline{2-2}
                                         & $U_{max}\approx0.966450101$,~ $U_{min}=-0.3435414897$                                         \\\hline  \hline                          
  \end{tabular}
  \centerline{
    \begin{minipage}{0.9\columnwidth}
    \caption{Hairy black hole and time-like geodesic parameters.}
    \label{tab1}
    \end{minipage}
    }
\end{table}

As we can see in Figure \ref{Potential}, apparently there is nothing new compared to the Schwarzschild case, however, in what follows we will present a new geodesic behavior given by the presence of the scalar field.

\begin{figure}
	\begin{subfigure}{.5\textwidth}
	  \centering
	  \includegraphics[width=.8\linewidth]{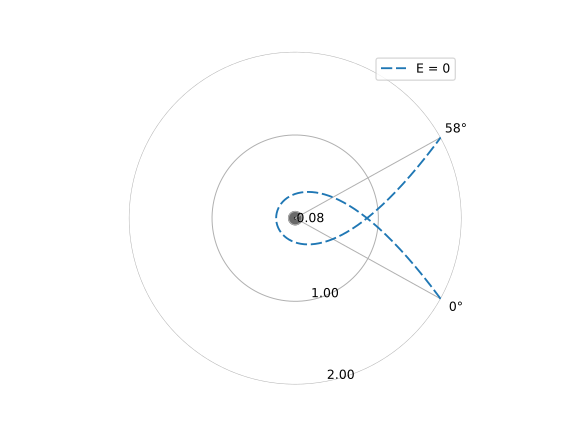}
	  \caption{$r_0=2~[AU]$ and $0<\varphi<2.5\pi$}
	  \label{fig6:1}
	\end{subfigure}%
	\begin{subfigure}{.5\textwidth}
	  \centering
	  \includegraphics[width=.8\linewidth]{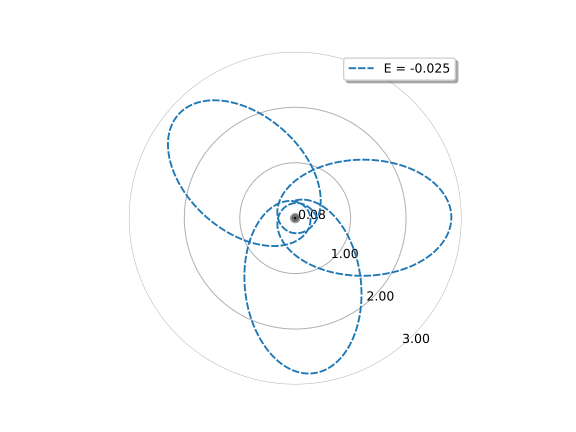}
	  \caption{$r_0=0.4~[AU]$  and $0<\varphi<8.5\pi$}
	  \label{fig6:2}
	\end{subfigure}
	\caption{\small\textbf{Negative Branch (time-like)} In order to compare we plot the Schwarzschild black hole horizon $r_h=\dfrac{2G_{N}M}{c^{2}}=0.08~AU$ (outer grey circle) and the hairy black hole horizon $\sqrt{\Omega(x_{h})}=0.009~AU$ (inner dark circle). The other constants are fixed to $\bar{\mathcal{J}}=2.6\times 10^{-6}~yr$, $\alpha=1$, $\nu=1.52$,   $E=\bar{\mathcal{E}}^{2}-1$.}
	\label{fig6}
\end{figure}

\begin{figure}
	\begin{subfigure}{.5\textwidth}
	  \centering
	  \includegraphics[width=.8\linewidth]{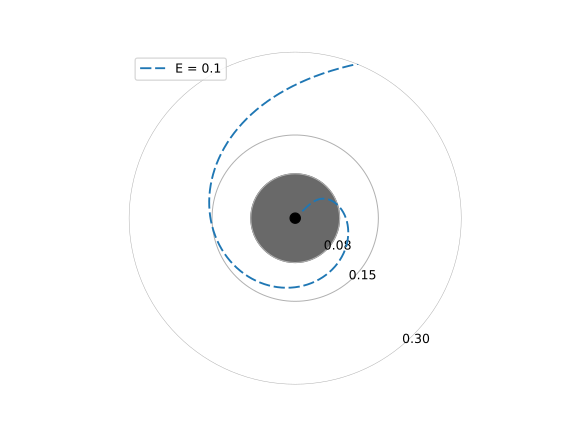}
	  \caption{$r_0=2~[AU]$  and $0<\varphi<2.5\pi$}
	  \label{fig7:1}
	\end{subfigure}%
	\begin{subfigure}{.5\textwidth}
	  \centering
	  \includegraphics[width=.8\linewidth]{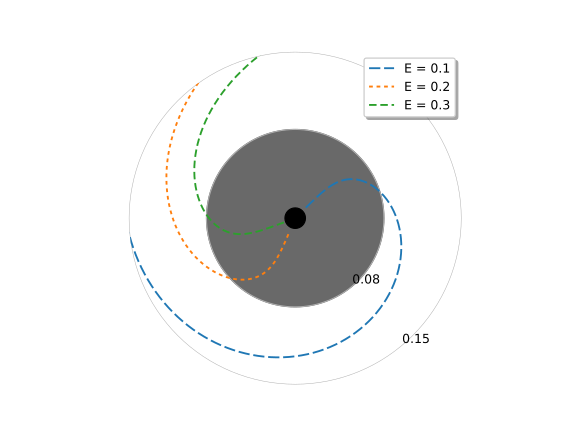}
	  \caption{$r_0=2~[AU]$ and $0<\varphi<2.5\pi$}
	  \label{fig7:2}
	\end{subfigure}
	\caption{\small\textbf{Negative Branch (time-like)} The hairy black hole is represented by the black disk of radius $\sqrt{\Omega(x_{h})}=0.009~AU$, while the Schwarzschild black hole, by the grey disk of radius  $r_h=\dfrac{2G_{N}M}{c^{2}}=0.08~AU$.
    The constants and parameters are fixed to $\bar{\mathcal{J}}=2.6\times 10^{-6}~yr$, $\alpha=1$, $\nu=1.52$, $\eta=12.527~AU^{-1}$, $E=\bar{\mathcal{E}}^{2}-1$. }
	\label{fig:7}
\end{figure}

%%%%%%%%%%%%%%%%%%%%%%%%%%%%%%%
\begin{figure}
	\begin{subfigure}{.5\textwidth}
	  \centering
	  \includegraphics[width=.8\linewidth]{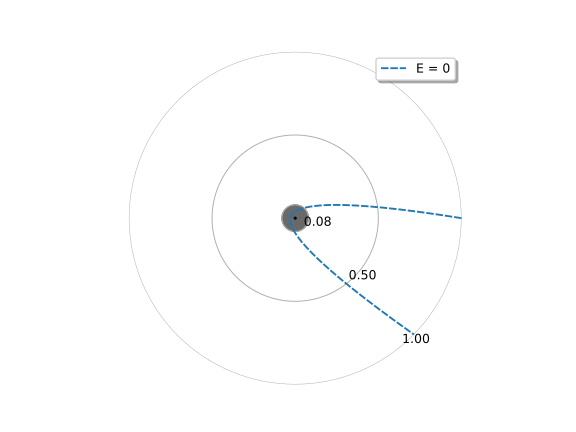}
	  \caption{$r_0=1~[AU]$ and $0<\varphi<1.7\pi$}
	  \label{fig8:1}
	\end{subfigure}%
	\begin{subfigure}{.5\textwidth}
	  \centering
	  \includegraphics[width=.8\linewidth]{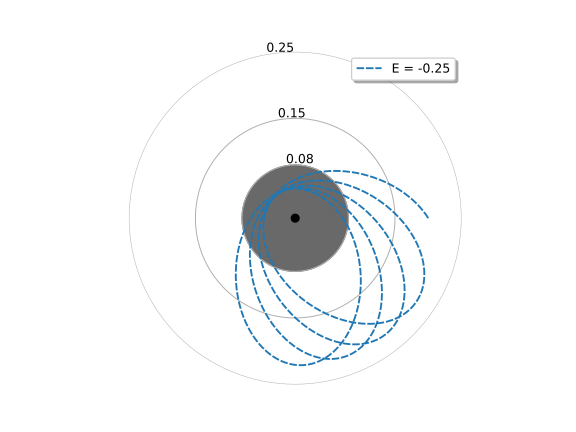}
	  \caption{$r_0=0.2~[AU]$  and $0<\varphi<8.5\pi$}
	  \label{fig8:2}
	\end{subfigure}
	\caption{\small\textbf{Positive Branch (time-like)} In both figures the hairy black hole is represented by the black disk of radius $\sqrt{\Omega(x_{h})}=0.006~AU$, while the Schwarzschild black hole, by the grey disk of radius  $r_h=2G_{N}M/c^{2}=0.08~AU$. The constants and parameters are fixed to $\bar{\mathcal{J}}=7\times 10^{-7}~yr$, $\alpha=-40$, $MG_{N}/c^2=0.04$, $\nu=1.76$, $E=\bar{\mathcal{E}}^{2}-1$.}
	\label{fig8}
\end{figure}

%%%%%%%%%%%%%%%%%%%%%%%%%%%%%%%
\begin{figure}
	\begin{subfigure}{.5\textwidth}
	  \centering
	  \includegraphics[width=.8\linewidth]{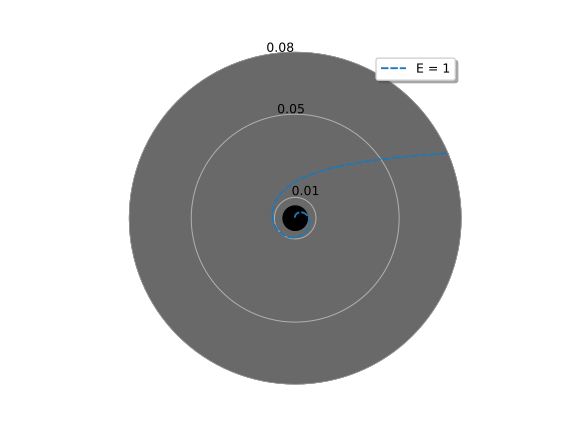}
	  \caption{$r_0=0.25~[AU]$  and $0<\varphi<2.5\pi$}
	  \label{fig9:1}
	\end{subfigure}%
	\begin{subfigure}{.5\textwidth}
	  \centering
	  \includegraphics[width=.8\linewidth]{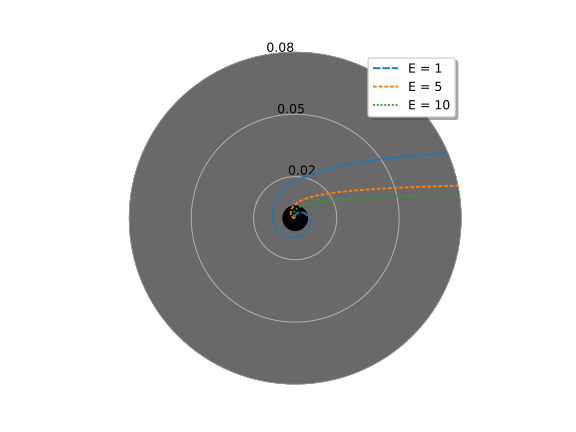}
	  \caption{$r_0=2~[AU]$ and $0<\varphi<2.5\pi$}
	  \label{fig9:2}
	\end{subfigure}
	\caption{\small\textbf{Positive Branch (time-like)}  The hairy black hole is represented by the black disk of radius $\sqrt{\Omega(x_{h})}=0.006~AU$, while the Schwarzschild black hole, by the grey disk of radius  $r_h=2G_{N}M/c^{2}=0.08~AU$. The other constants are fixed to $\bar{\mathcal{J}}=7\times 10^{-7}~yr$, $\alpha=-40$, $MG_{N}/c^2=0.04$, $\nu=1.76$, $E=\bar{\mathcal{E}}^{2}-1$.}
	\label{fig:9}
\end{figure}

%%%%%%%%%%%%%%%%%%%%%%%%%%%%%%
\section{Hairy null geodesics}
\label{sec:5}
%%%%%%%%%%%%%%%%%%%%%%%%%%%%%%
%
The parametric equation for null geodesics\footnote{Here the Killing vectors give us
  \begin{equation}
    \frac{dt}{d\lambda}=
    \frac{1}{bc~\Omega(x)f(x)}, \qquad \frac{d\varphi}{d\lambda}=\frac{c}{\Omega(x)}
  \end{equation}} will be shown in \cite{david1511}, and it can be easily calculated 
considering $ds^{2}=0$. 
We replace the proper time $\tau$ by an affine parameter $\lambda$ and scale it as $\lambda\rightarrow\lambda/\bar{\mathcal{J}}$
\begin{equation}
  \eta^2\Omega^{2}\qty(\frac{dx}{d\lambda})^{2}+\mathcal{V}(x)=\frac{c^2}{b^2}, \qquad
  \mathcal{V}(x)=f(x) c^{2}
  \label{geonull2}
\end{equation}
where the effective potential $\mathcal{V}(x,\alpha,M,\nu)$ is given by
\begin{equation}
  \mathcal{V}(x,\alpha,M,\nu)=c^{2}f(x)=\alpha c^{2}
  \biggl{[}\frac{1}{\nu^{2}-4}
  -\frac{x^{2}}{\nu^{2}%
  }\biggl{(}1+\frac{x^{-\nu}}{\nu-2}-\frac{x^\nu}{\nu+2}%
  \biggr{)}\biggr{]}+\frac{xc^2}{\Omega(x,M)}
  \label{eff1}
\end{equation}
The null geodesics are completely determined by the impact parameter 
$b^2\equiv(c\bar{\mathcal{J}})^2/\bar{\mathcal{E}}^2$. In order to integrate numerically the geodesic equation we need consider the first order orbital equation for $x(\varphi)$
\begin{equation}
  \qty(\frac{dx}{d\varphi})^{2}=\frac{1}{\eta^{2}}\qty[\frac{1}{b^{2}}-f(x)]
  \label{geonull3}
\end{equation}
Near to boundary we can integrate (\ref{geonull3}) and using $x=1-1/\eta r$
we get $r\varphi=b$,
which give us the intuitive definition of the impact parameter. 
Taking the derivative with respect to $\varphi$ in (\ref{geonull3}) we can get the second order equation
\begin{equation}
  \dv[2]{x}{\varphi}+\frac{1}{2\eta^{2}}\frac{df(x)}{dx}=0
\end{equation}
The extreme points
of the potential
where the location is $x_{0}$
\begin{equation}
  \mathcal{V}(x_{0})=\frac{c^{2}}{b_{0}^{2}}, \qquad \frac{d\mathcal{V}(x_{0})}{dx}=0
  ~~\Rightarrow ~~
  x_{0}^{\nu}=\frac{\alpha+\eta^{2}(2-\nu)}{\alpha+\eta^{2}(2+\nu)}
  \label{ec11}
\end{equation}
and the radius of the minimal unstable circular orbit is $r_{0}=\sqrt{\Omega(x_{0})}$, whit its respective critical impact parameter $b_{0}$
\begin{equation}
  \begin{split}
    r_{0} & =\pm \frac{1}{2\eta^{3}}[\alpha+\eta^{2}(2-\nu)]^{\frac{\nu-1}{2\nu}}[\alpha+\eta^{2}(2+\nu)]^{\frac{\nu+1}{2\nu}} \\
    x_{0} & <1~(+),  ~~ x_{0}>1~(-)
    \label{runs}
  \end{split}
\end{equation}
\begin{equation}
  \mathcal{V}(x_{0})=\frac{c^{2}}{b_{0}^{2}} ~~ \Rightarrow ~~
  b_{0}=\frac{1}{\sqrt{f(x_{0})}}
\end{equation}
Then
\begin{itemize}
  \item The light can be deflected if~~ $\Rightarrow ~~\frac{c^{2}}{b^{2}}<\mathcal{V}(x_{0})$
  \item The light get down to black hole if~~ $\Rightarrow~~ \frac{c^{2}}{b^{2}}>\mathcal{V}(x_{0})$
  \item The critical impact parameter is defined like a place in which the massless particles are trapped in an unstable circular orbit(ISCO)~~
        $\Rightarrow~~
          \mathcal{V}(x_{0})=\frac{c^{2}}{b_{0}^{2}}$
  \item In the no-hair limit $\nu=1$ the ISCO radius and the critical impact parameter are $r_{0}=3MG_{N}/c^{2}=0.12\, AU,~b_{0}=3\sqrt{3}MG_{N}/c^{2}=0.207846\, AU$.
\end{itemize}
\begin{table}[ht]
  \centering
  \begin{tabular}{c c}
    \hline \hline
    \multirow{5}*{Negative Branch} & $x<1$, $\alpha=1 AU^{-2}$, $\nu=1.52$                 \\ \cline{2-2}
                                   & $\eta=12.52655373~AU^{-1}$,~ $G_{N}M/c^{2}=0.04\,AU$  \\ \cline{2-2}
                                   & $x_{0}=0.271628$,~ $r_{0}=0.1003~AU$                  \\ \cline{2-2}
                                   & $b_0=0.192949\,AU$                                    \\ \cline{2-2}
                                   & $\mathcal{V}(x_{0})=1.07423\times 10^{11}\,yr^{-2}$   \\ \hline
    \multirow{5}*{Positive Branch} & $x>1$, $\alpha=-40~AU^{-2}$, $\nu=1.76$               \\ \cline{2-2}
                                   & $\eta=3.252719443~AU^{-1}$,~ $G_{N}M/c^{2}=0.04\, AU$ \\ \cline{2-2}
                                   & $x_{0}=18.5943$,~ $r_{0}=0.0096397 ~AU$               \\ \cline{2-2}
                                   & $b_0=0.0327298\,AU$                                   \\ \cline{2-2}
                                   & $\mathcal{V}(x_{0})=3.73333\times 10^{12}\,yr^{-2}$   \\ \hline \hline
  \end{tabular}
  \caption{Hairy black hole and null geodesic parameters.}
  \label{tab2}
\end{table}

\begin{figure}
	\begin{subfigure}{.5\textwidth}
	  \centering
	  \includegraphics[width=.8\linewidth]{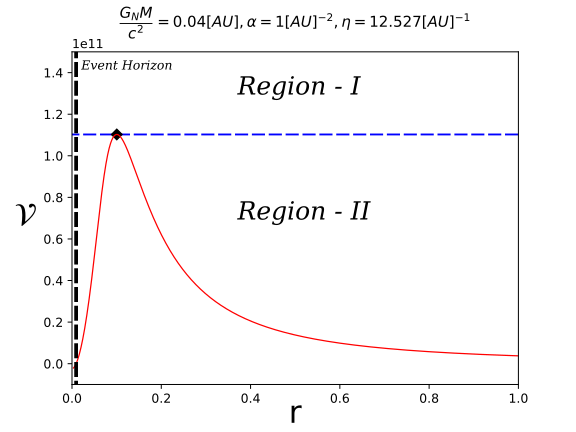}
	  \caption{Negative branch}
	  \label{fig10a}
	\end{subfigure}%
	\begin{subfigure}{.5\textwidth}
	  \centering
	  \includegraphics[width=.8\linewidth]{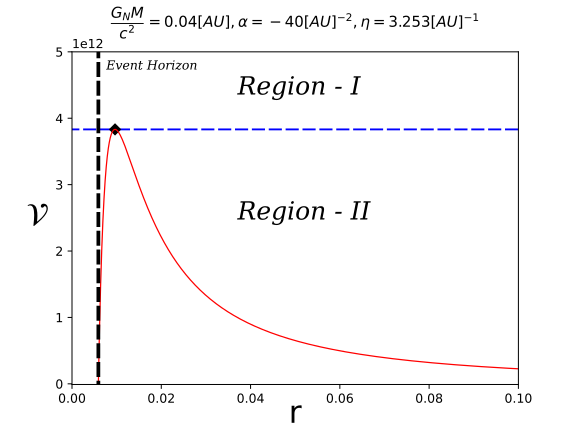}
	  \caption{Positive branch}
	  \label{fig10b}
	\end{subfigure}
	\caption{\small We have the positive-branch (\subref{fig10b}) and negative-branch (\subref{fig10a}) of the effective potential (\ref{eff1}). The parameters are described in table (\ref{tab2}).}
	\label{pote1}
\end{figure}

%
%%%%%%%%%%%%%%%%%%%%%%%%%%%%%%%%%%%%%%%%%
%

\begin{figure}
\begin{subfigure}{.5\textwidth}
	\centering
	\includegraphics[width=.8\linewidth]{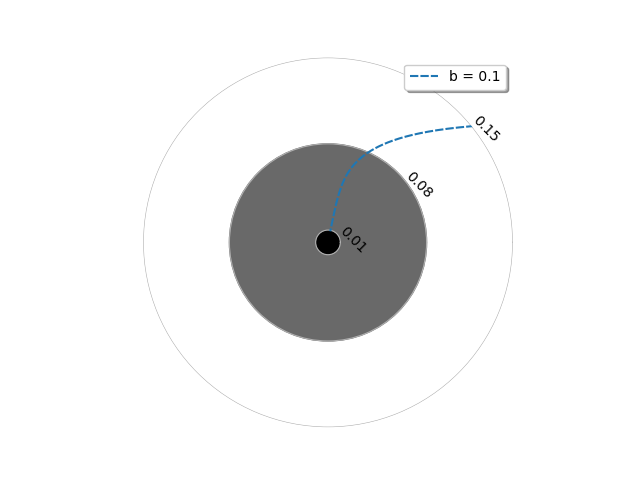}
	\caption{$r_0=1~[AU]$  and $0.1<\varphi<1.2\pi$}
	\label{fig11a}
\end{subfigure}%
\begin{subfigure}{.5\textwidth}
	\centering
	\includegraphics[width=.8\linewidth]{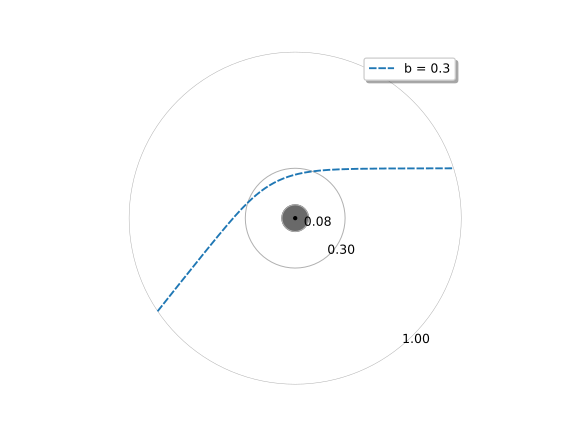}
	\caption{$r_0=1~[AU]$ and $0.29<\varphi<1.2\pi$}
	\label{fig11b}
\end{subfigure}
\caption{\small\textbf{Negative Branch (null)}. The Schwarzschild black hole horizon is $r_h=\dfrac{2G_{N}M}{c^{2}}$ (grey circle) and hairy horizon $\sqrt{\Omega(x_{h})}$ (black circle). The other constants are fixed to $\alpha=1$, $MG_{N}/c^{2}=0.04$, $\nu=1.52$.}
\label{fig:11}
\end{figure}

\noindent
From Figure \ref{fig:11} we can conclude
\begin{itemize}
  \item [\ref{fig11a}] According to table (\ref{tab2}) the null geodesic is defined in the region-I, see (\ref{fig10a}), where $\frac{c^2}{b^2}=4.10496\cdot 10^{11}~yr^{-2}>\mathcal{V}(x_{0})$. A null-particle fall to the black hole, we observe that the orbit is perpendicular to the hairy black hole horizon. The backreaction in the dense hair region cause this anomalous orbit.\\
  \item [\ref{fig11b}] According to table (\ref{tab2}) the null geodesic is defined in the region-II, see (\ref{fig10a}), where $\frac{c^2}{b^2}=0.45611\cdot 10^{11}~yr^{-2}<\mathcal{V}(x_{0})$. A null-particle is deflected by the hairy black hole
\end{itemize}

\begin{figure}
\begin{subfigure}{.5\textwidth}
	\centering
	\includegraphics[width=.8\linewidth]{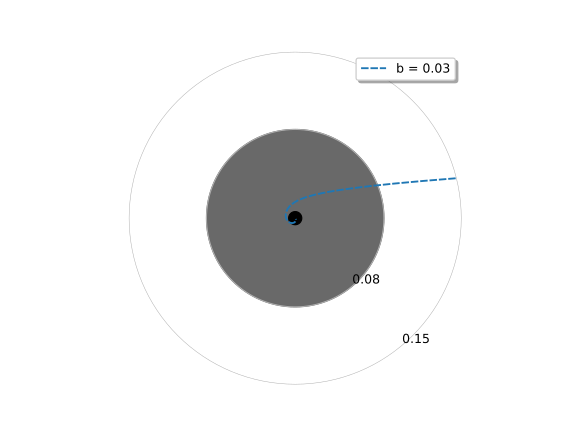}
	\caption{$r_0=1~[AU]$  and $0.03<\varphi<1.7\pi$}
	\label{fig12a}
\end{subfigure}%
\begin{subfigure}{.5\textwidth}
	\centering
	\includegraphics[width=.8\linewidth]{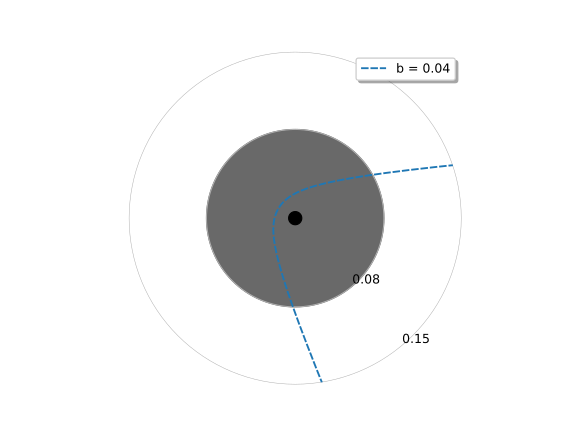}
	\caption{$r_0=1~[AU]$ and $0.04<\varphi<1.6\pi$}
	\label{fig12b}
\end{subfigure}
\caption{\small\textbf{Positive Branch (null)}
The Schwarzschild black hole horizon is $r_h=\dfrac{2GM}{c^{2}}$ (grey disk contour) and hairy horizon $\sqrt{\Omega(x_{h})}$ (black disk contour). The other constants are fixed to $\alpha=-40$, $MG_{N}/c^{2}=0.04$, $\nu=1.76$.}
\label{fig:12}
\end{figure}

\noindent
From Figure \ref{fig:12} we can conclude
\begin{itemize}
  \item [\ref{fig12a}] According to table (\ref{tab2}) the null geodesic is defined in region-I, see (\ref{fig10b}), where $\frac{c^2}{b^2}=4.56107\cdot 10^{12}~yr^{-2}>\mathcal{V}(x_{0})$. A null-particle fall to the black hole, we observe that the orbit is not perpendicular to the hairy black hole horizon. That orbit is usual in Schwarzschild black hole space-time. \\
  \item [\ref{fig12b}] According to table (\ref{tab2}) the null geodesic is defined in region-II, see (\ref{fig10b}), where $\frac{c^2}{b^2}=2.56560\cdot 10^{12}~yr^{-2}<\mathcal{V}(x_{0})$. A null-particle is highly deflected. Indeed, this light deviation is impossible to find in Schwarzschild black hole space-time.
\end{itemize}

%%%%%%%%%%%%%%%%%%%%%%%%%%%%%%%%%%%%%%%
\section{Hairy near horizon geodesics}
\label{sec:6}
%%%%%%%%%%%%%%%%%%%%%%%%%%%%%%%%%%%%%%%

The goal of the present section is to explain the anomalous infalling time-like or space-like geodesics shown at Figure \ref{fig:7} and \ref{fig:11}. In those figures the geodesics apparently go inside the black hole orthogonal to the horizon surface, here we verify that this is actually true.\\

Considering the solution of the hairy black hole 
given in section \ref{sec:2}, the localization of the horizon $x_{h}$
is such that $f(x_{h})=0$ and the near horizon
geometry can be constructed under the following change $x=x_{h}+\epsilon$, giving us
\begin{equation}
  \begin{split}
    f\Omega\vert_{x_{h}+\epsilon} & \approx(x-x_{h})\Omega(x_{h})f^{'}(x_{h}), \\ 
    \frac{f}{\Omega}\br\vert_{x_{h}+\epsilon} & \approx (x-x_{h})\frac{f^{'}(x_{h})}{\Omega(x_{h})}
  \end{split}
\end{equation}
replacing in the hairy metric (\ref{Ansatz}) 
\begin{equation}
  ds^{2}=-(x-x_{h})\Omega(x_{h})f^{'}(x_{h})c^{2}dt^{2}+\frac{\eta^{2}dx^{2}}{(x-x_{h})\frac{f^{'}(x_{h})}{\Omega(x_{h})}}
  +\Omega(x_{h})(d\theta^{2}+\sin^{2}{\theta}d\varphi^{2})
  \label{hairynear}
\end{equation}
taking the usual transformation to radial part
\begin{equation}
  d\rho^{2}=\frac{\eta^{2}dx^{2}}{(x-x_{h})\frac{f^{'}(x_{h})}{\Omega(x_{h})}}~\Rightarrow ~ \rho^{2}=4\eta^{2}\frac{\Omega(x_{h})}{f^{'}(x_{h})}(x-x_{h})
\end{equation}
and for the temporal coordinate $t_{R}=\frac{f^{'}(x_{h})}{2\eta}t$, we get the Rindler geometry
\begin{equation}
  ds^{2}=-\rho^{2}dt_{R}^{2}+d\rho^{2}+\frac{\Omega(x_{h})}{l^{2}}(d\theta^{2}+\sin^{2}{\theta}d\varphi^{2})
\end{equation}
%
%%%%%%%%%%%%%%%%%%%%%%%%%
\subsection{Time-like near horizon geodesics}
%%%%%%%%%%%%%%%%%%%%%%%%%
%
The near horizon geometry of (\ref{Ansatz})  is described by the metric
\begin{equation}
  \begin{split}
    ds^{2} & =\Omega(x_{h})\qty(-F(x)c^{2}dt^{2}+\frac{\eta^{2}dx^{2}}{F(x)}+d\theta^{2}+\sin{\theta}^{2}d\varphi^{2})  \\
    F(x) & =(x-x_{h})f^{\prime}(x_{h})
    \label{near1}
  \end{split}
\end{equation}
%
%where
%
%\begin{equation}
%\Omega(x)~\Rightarrow~\Omega(x_{h}), \qquad
%f(x)~\Rightarrow~
%F(x)=(x-x_{h})f^{\prime}(x_{h})
%\end{equation}
%
The Killing equations give us the following conserved quantities
\begin{equation}
  \dot{t}=\frac{\bar{\mathcal{E}}}{\Omega(x_{h})F(x)}, \qquad
  \dot{\varphi}=\frac{\bar{\mathcal{J}}c^{2}}{{\Omega(x_{h})}}
  \label{near2}
\end{equation}
And the parametric equation is
\begin{equation}
  \begin{split}
    \bar{\mathcal{E}}^{2}-1
    =\left(\frac{\eta\Omega(x_{h})}{c}\right)^2
    \qty(\dv{x}{\tau})^{2}+U_{\text{eff}}(x) \\
    U_{\text{eff}}(x)=\Omega(x_{h})F(x)\left(1+\frac{\bar{\mathcal{J}^{2}}c^{2}}{\Omega(x_{h})}\right)-1
    \label{pothairy11}
  \end{split}
\end{equation}
Considering the following chain-rule: $\dot{x}=\dv{x}{\varphi}\dot{\varphi}$, we can get the orbital equation with $E\equiv\bar{\mathcal{E}}^{2}-1$
\begin{equation}
  E=\eta^{2}\bar{\mathcal{J}}^{2}c^{2}\qty(\dv{x}{\varphi})^{2}+U_{eff}(x)
\end{equation}
taking the derivative with respect to $\varphi$
\begin{equation}
  \begin{split}
    \frac{d^{2}x}{d\varphi^{2}}+ & H(x_{h},\nu,\bar{\mathcal{J}},\eta)=0, \\
    H(x_{h},\nu,\bar{\mathcal{J}},\eta) & =\frac{f^{\prime}(x_{h})}{2(\eta\bar{\mathcal{J}}c)^{2}}\qty[\Omega(x_{h})+(\bar{\mathcal{J}}c)^{2}]
    \label{TMgeo1}
  \end{split}
\end{equation}
the solution is
\begin{equation}
  x(\varphi)=-\frac{H(x_{h},\nu,\bar{\mathcal{J}},\eta)}{2}\varphi^{2}+c_{1}\varphi+c_{2}
\end{equation}
The initial conditions are, see the example of Schwarzschild case in the \ref{nearSch},
\begin{equation}
  x(0)=x_{h}~\Rightarrow~c_{2}=x_{h}, \qquad
  \qty(\dv{x}{\varphi})_{x_{h}}=\frac{\bar{\mathcal{E}}}{\eta\bar{\mathcal{J}}c}=c_{1}
  \label{icnear}
\end{equation}
the near-horizon solution for time-like geodesic, with \eqref{icnear}, is 
\begin{equation}
  x(\varphi)=-\frac{H(x_{h},\nu,\bar{\mathcal{J}},\eta)}{2}\varphi^{2}+\frac{\bar{\mathcal{E}}}{\eta\bar{\mathcal{J}}c}\varphi+x_{h}
  \label{xnear1}
\end{equation}
and to plotting we consider: $r(\varphi)=\sqrt{\Omega(x(\varphi))}$
\subsection{Null-like near horizon geodesics}
Considering the near-horizon metric (\ref{near1}) and the change (\ref{near2}) we obtain the
following expressions for
conserved quantities
\begin{equation}
  \frac{dt}{d\xi}=
  \frac{1}{bc~\Omega(x_{h})F(x)}, \qquad \frac{d\varphi}{d\xi}=\frac{c}{\Omega(x_{h})}
\end{equation}
the first order orbital equation is
\begin{equation}
  \qty(\frac{dx}{d\varphi})^{2}=\frac{1}{\eta^{2}}\qty[\frac{1}{b^{2}}-F(x)]
\end{equation}
the second order and its solution is
\begin{equation}
  \dv[2]{x}{\varphi}+\frac{f^{\prime}(x_{h})}{2\eta^{2}}=0 ~\Rightarrow~ x(\varphi)=-\frac{f^{\prime}(x_{h})}{4\eta^{2}}\varphi^{2}+c_{1}\varphi+c_{2}
\end{equation}
The initial conditions are
\begin{equation}
  x(0)=x_{h}~\Rightarrow~c_{2}=x_{h}, \qquad
  \qty(\dv{x}{\varphi})_{x_{h}}=\frac{1}{\eta b}=c_{1}
\end{equation}
The near-horizon solution 
for time-like geodesic is, here we would like to highlight that $x(\varphi)=x(\varphi,\nu)$ depend on 
the hairy parameter,
\begin{equation}
  x(\varphi)=-\frac{f^{\prime}(x_{h})}{4\eta^{2}}\varphi^{2}+\frac{\varphi}{\eta b}+x_{h}, \qquad r(\varphi)=\sqrt{\Omega(x(\varphi))}
  \label{xnear2}
\end{equation}
the above solution is very similar to the Schwarzschild case \eqref{b10}, however the constants that multiply $\varphi^{2}$ and $\varphi$ depend on the hairy parameter $\nu$, which clearly changes the usual Schwarzschild behaviour.\\
\begin{figure}
\centering
\includegraphics[width=0.4\linewidth]{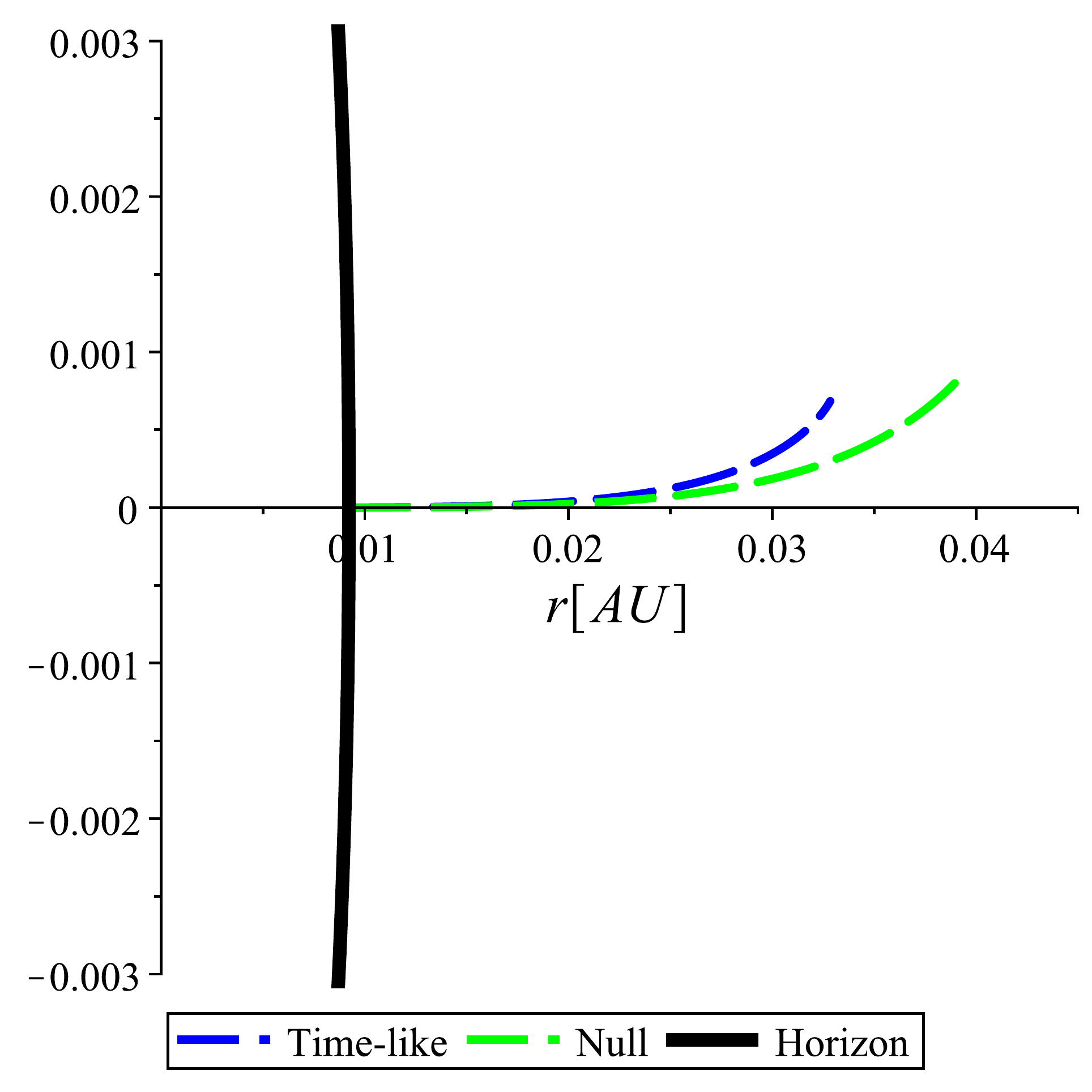}
\caption{\small\textbf{Hairy near horizon geodesics}.
Here we plot the near horizon geodesics of the Figure \ref{fig7:2} and Figure \ref{fig11b}, which are null and time-like respectively, using the equation (\ref{xnear1}) and (\ref{xnear2}). Here we consider the angular momenta per unit mas $\bar{\mathcal{J}}=2.6\times 10^{-6}$ and the mass of the black hole $G_{N}M/c^{2}=0.04$. For time-like geodesic we consider $E=0.1$ and for null case $b=0.1$, $E=0.1$. The black line define the horizon of the black hole of radius $r_{h}=0.0192~AU$.} 
\label{HairyNear}
\end{figure}

The Schwarzschild black hole has geodesics which fall inside of the black hole with some angle with respect to the tangent to horizon, see Figure \ref{SchNear}, while in the hairy case we showed that geodesics cross the horizon orthogonally.

\section{Discussion}
\label{sec:7}
Our first results are the existence of the critical mass for negative branch $0<x<1$ with $2<\nu$ and positive branch $1<x<\infty$ with $\nu>1$
\begin{equation}
  \begin{split}
    M_{cri}\equiv\frac{c^{2}}{2G_{N}}\qty(\frac{\nu-2}{\alpha})^{1/2} \\
    M_{cri}\equiv\frac{c^{2}(\nu-1)(\nu+2)}{6G_{N}\sqrt{-\alpha(\nu+2)}}
  \end{split}
\end{equation}
there is a black hole horizon if $M>M_{cri}$, see \ref{gttneg-A}, \ref{gttneg-B} and \ref{gttpo}, \ref{fig:po} 
then, the existence of the
scalar field $\phi(x)$ and the non-trivial potential $V(\phi)$, forces that if there is not enough mass, the black hole could implode leaving a naked singularity.
The existence of this critical mass could impose conditions
on the minimal mass in modeling of the accretion disk \cite{Cunha:2019hzj,Tian:2019yhn,Porth:2016rfi,Cunha:2018acu,PhysRevD.103.104047}

The most important result shown in section \ref{sec:5} is the anomalous changing of geodesics as particles enter into the dense hair region (grey disk), see \ref{fig:7} and \ref{fig:11}. Both types of geodesics go inside the hairy black hole orthogonal to the horizon surface, see \ref{HairyNear}. The great backreaction of the potential $V(\phi)$ in the dense hair region $\mathcal{D}\equiv 2MG_{N}/c^{2}-r_{h}$ is the main cause of anomalous changing of geodesics trajectories. It is anomalous because, for static black hole asymptotically flat solutions, there is no similar examples in the literature. To clarify this results, in section \ref{sec:6} we construct the near horizon geodesics for black holes described at Figure \ref{fig:7} and \ref{fig:11}, which results are shown in Figure \ref{HairyNear}. There are other hairy configurations, such as \ref{fig:9} and \ref{fig:12}, that do not present the anomalous changing of geodesics.
For certain values of the parameters of the theory $\alpha, \nu$ and energy we have geodesics topologically equivalent to those of Schwarzschild black hole and for another range of parameters $\alpha, \nu$ we obtain an anomalous behavior

The second future direction consists of studying how to take advantage of this anomalous effect in geodesics to know if it is a black hole with hair or not. A first proposal would be to launch a test satellite in a closed orbit, this will allow us to fix the parameters of the theory, $\alpha$ and $\nu$, and build the effective potential similar to \ref{fig5a} and \ref{fig5b}, then we look for an energy range in region I, in which geodesics have this anomalous behavior.
From a theoretical point of view we would like to find the exact range of the parameters of the theory in which this anomalous behavior exists.

In section \ref{sec:5} we get the radio $r_{0}$ of unstable circular orbit (ISCO), for negative branch $\phi<0$
\begin{equation}
r_{0} = \frac{1}{2\eta^{3}}[\alpha+\eta^{2}(2-\nu)]^{\frac{\nu-1}{2\nu}}[\alpha+\eta^{2}(2+\nu)]^{\frac{\nu+1}{2\nu}}
\end{equation}
and, $\phi>0$ for positive branch 
\begin{equation}
r_{0} =-\frac{1}{2\eta^{3}}[\alpha+\eta^{2}(2-\nu)]^{\frac{\nu-1}{2\nu}}[\alpha+\eta^{2}(2+\nu)]^{\frac{\nu+1}{2\nu}}
\end{equation}
which depend on the hairy parameters 
$\alpha, \nu$ and on the mass of the black hole $\eta(M,\alpha)$. Then, the shadow of the hairy black hole depend of $\alpha, \nu, M$. In a recent work \cite{PhysRevD.103.104047}, they have shown 
that we can constraint the  hairy parameters $\alpha, \nu, M$ based on size of the shadow of $M87^{*}$ black hole \cite{akiyama2019event,event2019first}. This will be one of the topics of a future work, in addition to modeling the accretion disk
%Finally the last interesting result is that the hairy configurations have shown a parabolic geodesics similar to Kepler case. This is an important
%observation because in the space-time of Schawarzschild black hole there is not 
%a parabolic geodesic, however the hair of the black hole allows existence

\section*{\normalsize Acknowledgments}
\vspace{-5pt}

Research of WC supported by Universidad Nacional de San Antonio Abad del Cusco.
The work of DC is supported by Pontificia Universidad Cat\'olica de Valpara\'\i so. The author GV-M acknowledges the receipt of the grant from the Abdus Salam International Centre for Theoretical Physics, Trieste, Italy.

\clearpage
\begin{appendices}
\renewcommand{\theequation}{\thesection\arabic{equation}}
%%%%%%%%%%%%%%%%%%%%%%%%%%%%%%%%%%%%%%%%%%%%%%%%%%%%%%%%%%%%%%
\section{Schwarzschild solution}
\label{ap:1}
%%%%%%%%%%%%%%%%%%%%%%%%%%%%%%%%%%%%%%%
%
\begin{equation}
  ds^{2}=-c^{2}N(r)dt^{2}+\frac{dr^{2}}{N(r)}+r^2\left(d\theta^{2}+\sin^{2}\theta d\varphi^{2}\right)
  \label{Sch1}
\end{equation}
\begin{equation}
  N(r)=1-\frac{r_{h}}{r}
\end{equation}
Here, $r_{h}$ is the horizon radius. The mass of the black hole is given by
\begin{equation}
  M=\frac{c^{2}r_{h}}{2G_{N}}
\end{equation}
%
%%%%%%%%%%%%%%%%%%%%%%%%%%%%%%%%
\subsection{Time-like geodesic}
\begin{figure}
\begin{subfigure}{.5\textwidth}
	\centering
	\includegraphics[width=.8\linewidth]{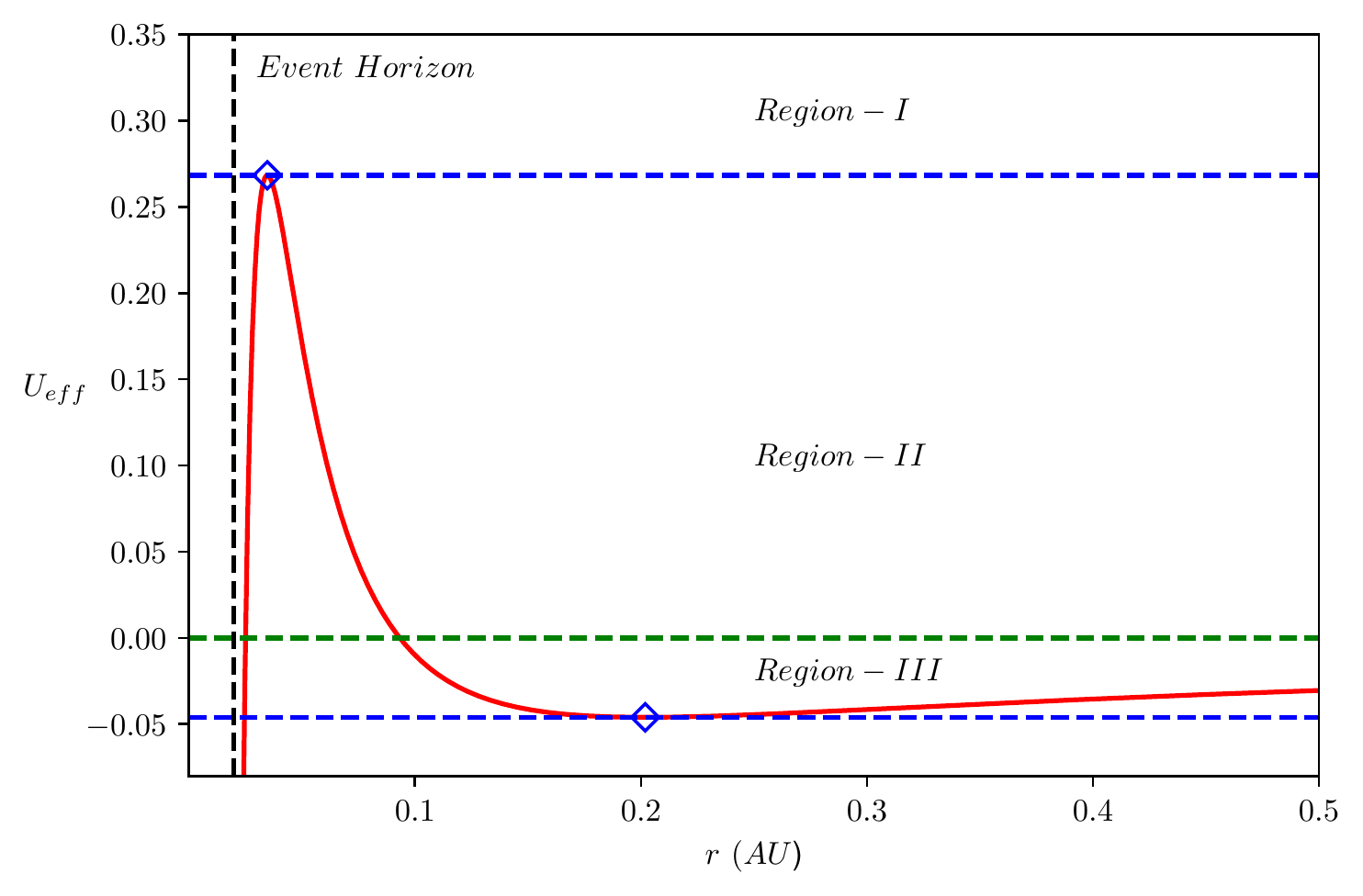}
	\caption{}
	\label{fig14a}
\end{subfigure}%
\begin{subfigure}{.5\textwidth}
	\centering
	\includegraphics[width=.8\linewidth]{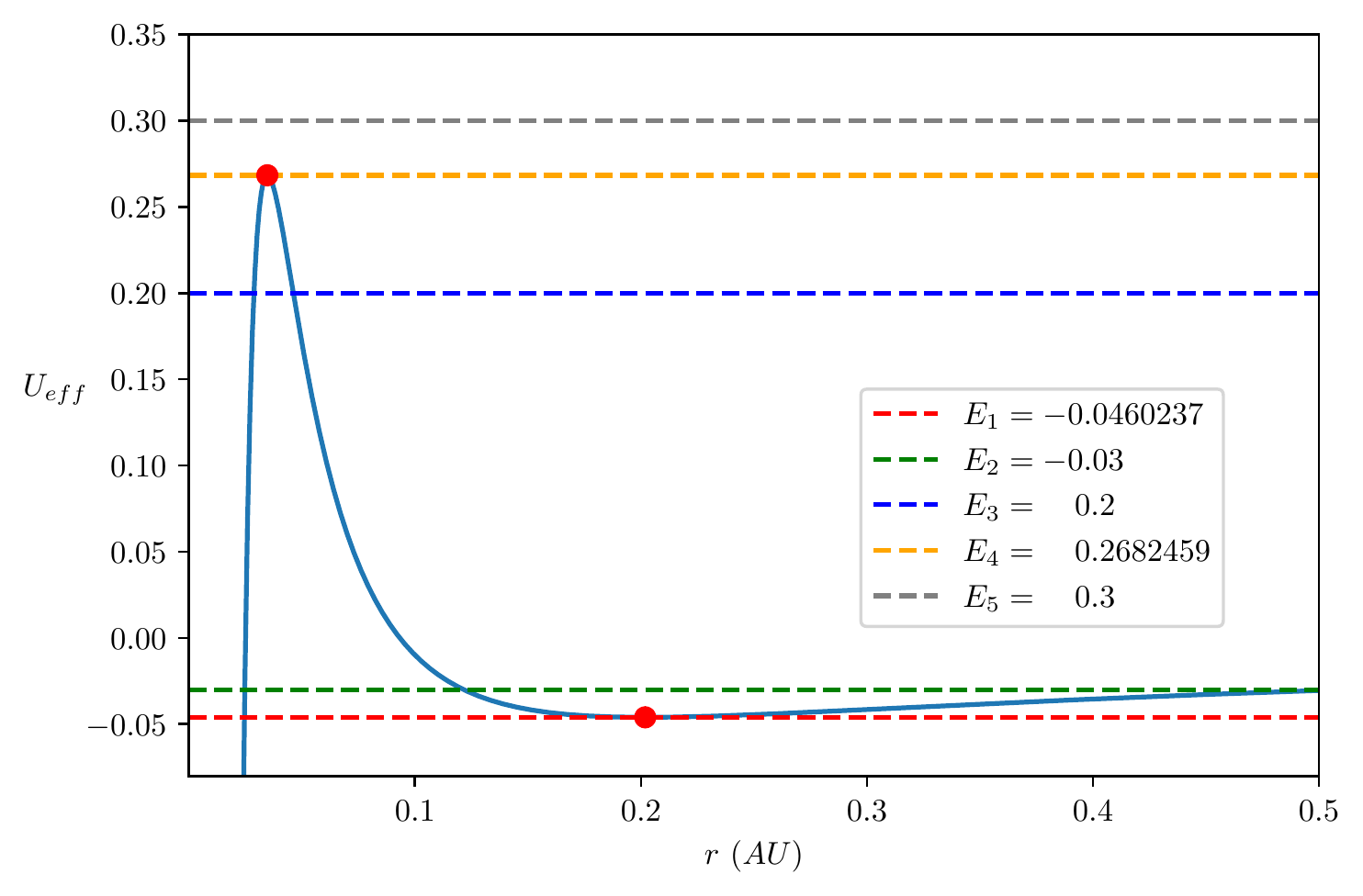}
	\caption{}
	\label{fig14b}
\end{subfigure}
\caption{\small (\subref{fig14a}):
Region I, Above $E=U^{max}_{eff}$. Region II: Between $E=0$ and $E=U^{max}_{eff}$. Region III: Between $E=U^{min}_{eff}$ and $E=0$. To make this plot we have considered $\bar{\mathcal{J}}^2=\frac{24M^2G^2_N}{c^6}$, with $M=10^6 M_\odot$.\\
(\subref{fig14b}): $E_1$ corresponds to the minimum of the effective potential. $E_2$ belongs to Region III. $E_3$ belongs to Region II. $E_4$ corresponds to the maximum of the effective potential. $E_5$ belongs to the region I.}
\label{fig:my_label}
\end{figure}

%%%%%%%%%%%%%%%%%%%%%%%%%%%%%%%%
\label{apendice1}
Here $ds^{2}=-c^{2}d\tau^{2}$, where $\tau$ is proper time. In addition, setting the geodesic at the equatorial plane $\qty(\theta=\pi/2)$ due to the rotational isometry we obtain
\begin{equation}
  - c^2
  =-N(r)c^{2}\dot{t}^{2}+\frac{\dot{r}^{2}}{N(r)}+r^{2}\dot{\varphi}^{2}
  \label{ec1}
\end{equation}
The conserved quantities along the isometry orbits generated by the Killing vectors $\xi_{(t)}=\partial_t$ and $\xi_{(\varphi)}=\partial_\varphi$ are given by $\bar{\mathcal{E}}$ (dimensionless) and $\bar{\mathcal{J}}~(yr)$: $dt/d\tau=\bar{\mathcal{E}}/N(r)$ and 
$d\varphi/d\tau=\bar{\mathcal{J}}c^{2}/r^{2}$. Here $\xi_{(t)}=(c,0,0,0)$ and $\xi_{(\varphi)}=(0,0,0,1)$, replacing in (\ref{ec1}) we can get the first order orbital equation. It describes the radial motion of a test body with energy $E=\bar{\mathcal{E}}^2-1$ in the effective potential, see Figure \ref{fig:my_label}
\begin{equation}
  E=\frac{\dot{r}^{2}}{c^{2}}+U_{\text{eff}}(r), \qquad 
  U_{\text{eff}}(r)=N(r)\left(1+\frac{\bar{\mathcal{J}}^{2}c^{2}}{r^{2}}\right)-1
  \label{poo}
\end{equation}
The polar equation can be constructed 
considering the chain-rule $\frac{dr}{d\tau}=\frac{dr}{d\varphi}\frac{d\varphi}{d\tau}$, from that we have
\begin{equation}
  \qty(\frac{dr}{d\varphi})^{2}  =
  \frac{r^{4}}{\bar{\mathcal{J}}^{2}c^{2}}(E-U_{eff}), \qquad
  \Delta\varphi  =2\qty|\int_{r_{p}}^{r_{a}}\frac{c\bar{\mathcal{J}}dr}{r^{2}\sqrt{E-U_{eff}(r)}}|-2\pi
  \label{devi1}
\end{equation}
the left-hand equation describe the orbit in polar coordinates, and the right-hand equation is the precession of the orbits closed for each revolution, where $r_{a}, r_{p}$ can be solved from $\eval{\frac{dr}{d\varphi}}_{r_{a},r_{p}}=0$ or $E=U_{eff}(r_{a},r_{p})$. The typical second order orbital equation
can be get considering the following change of variables $u(\varphi)=1/r(\varphi)$ and $du/d\varphi=-u^{2}dr/d\varphi$
\begin{equation}
  \frac{d^{2}u}{d\varphi^{2}}+u-\dfrac{3r_h}{2}u^2=\dfrac{1}{\lambda}, \qquad \lambda=\frac{2c^{2}\bar{\mathcal{J}}^{2}}{r_{h}}=\frac{2\mathcal{J}^{2}}{m^2c^2r_{h}}
  \label{geosh}
\end{equation}
We use the Runge-Kutta method in order to solve it

\clearpage
\begin{center}
  \textbf{Orbits}
\end{center}
Here we present the plots corresponding to each energy level presented in Figure \ref{fig:my_label}.

\begin{figure}
\begin{subfigure}{.5\textwidth}
	\centering
	\includegraphics[width=.8\linewidth]{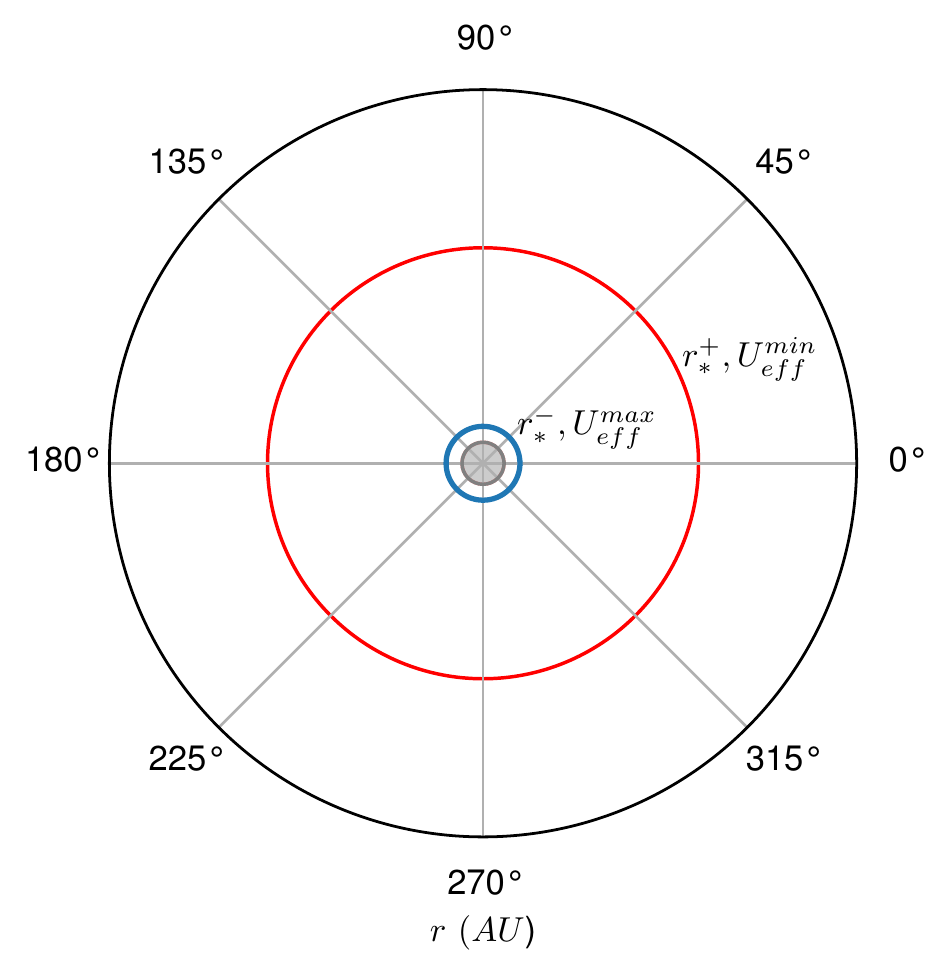}
	\caption{$r_h$, the stable circular orbit (red), \\and the unstable circular orbit (blue).}
	\label{time14:1}
\end{subfigure}%
\begin{subfigure}{.5\textwidth}
	\centering
	\includegraphics[width=.8\linewidth]{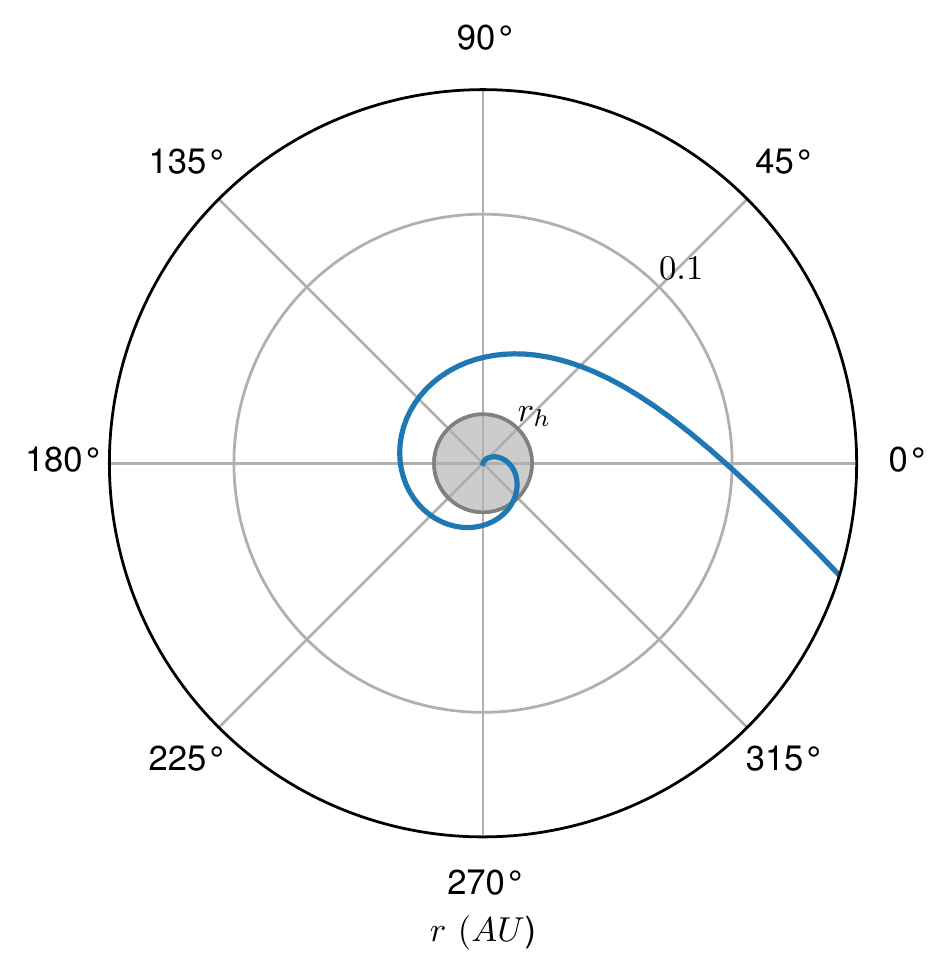}
	\caption{$r_h$ and the orbit (blue).}
	\label{time14:2}
\end{subfigure}
\caption{\small\textbf{ (Time-like)}\\(\subref{time14:1}) $E_1=U_{eff}^{min}=U_{eff}(r_{*}^+)$, initial conditions: $r_0=r_{*}^+$ and $\dot{r}_0=0$. $E_4=U_{eff}^{max}=U_{eff}(r_{*}^-)$, initial conditions:  $r_0=r_{*}^-$ and $\dot{r}_0=0$. (\subref{time14:2}) $E=E_5=0.3$, initial conditions: $r_0\rightarrow\infty$ and $\dot{r}_0=\sqrt{Ec^2}$. }
\label{time14}
\end{figure}

\begin{figure}
\begin{subfigure}{.5\textwidth}
	\centering
	\includegraphics[width=.8\linewidth]{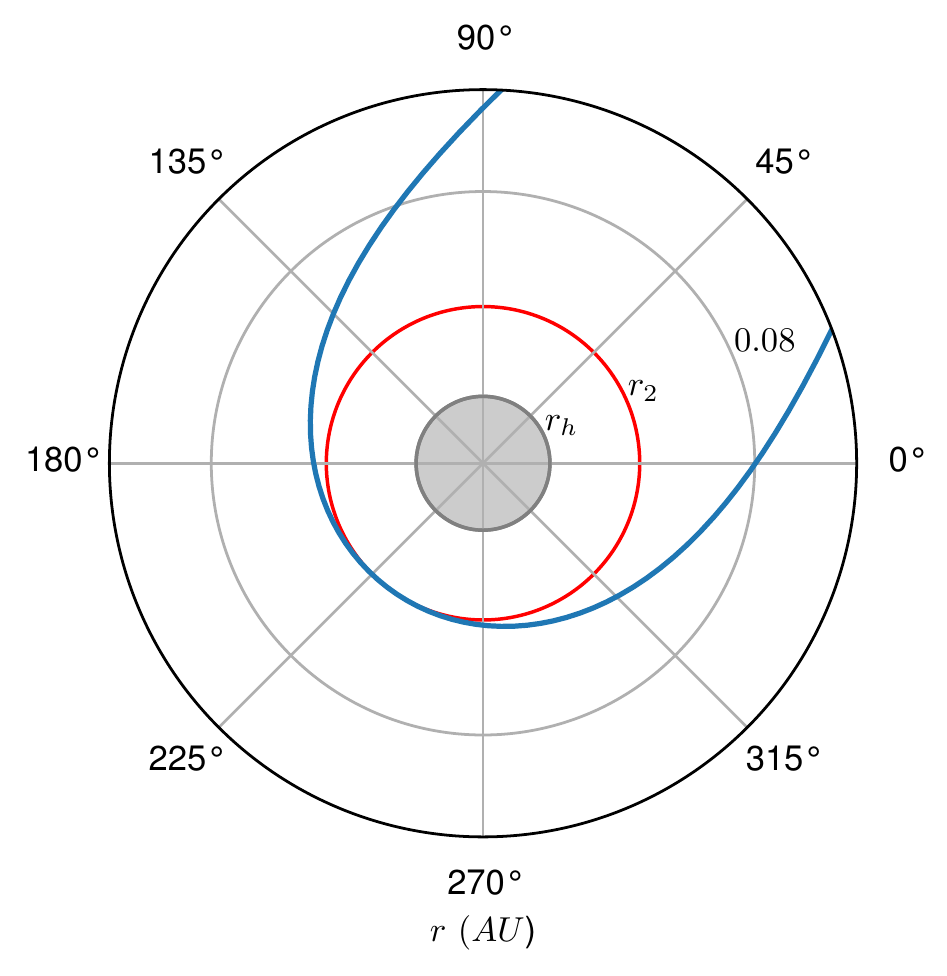}
	\caption{Close-up view: $r_h$, $r_2$ (red) and\\the orbit (blue). Where $r_2$ indicates the radius of the minimum distance from the center of the black hole to the orbit.}
	\label{fig:appx:1}
\end{subfigure}%
\begin{subfigure}{.5\textwidth}
	\centering
	\includegraphics[width=.8\linewidth]{fig/figE3-lejos.pdf}
	\caption{Distant view: $r_h$, $r_2$ (red) and\\the orbit (blue)}
	\label{fig:appx:2}
\end{subfigure}
\caption{\small\textbf{ (Time-like)}\\$E=E_3=0.2$, initial conditions: $r_0\rightarrow\infty$ and $\dot{r}_0=\sqrt{Ec^2}$. Where, $r_2=0.04613484978865398\,\,AU$ indicates the radius of the minimum distance from the center of the black hole to the orbit.}
\label{fig:appx}
\end{figure}

\clearpage
%%%%%%%%%%%%%%%%%%%%%%%
\subsection{Null geodesic}
%%%%%%%%%%%%%%%%%%%%%%%
%
Here we consider an affine parameter $\xi$ and we scale it as
$\xi\rightarrow \xi/\bar{\mathcal{J}}$
\begin{equation}
  \frac{dt}{d\xi}=
  \frac{\bar{\mathcal{E}}}{\bar{\mathcal{J}}N(r)}, \qquad
  \frac{d\varphi}{d\xi}=\frac{c^{2}}{r^{2}}
\end{equation}
replacing in (\ref{ec1}) at $ds^{2}=0$ we get
\begin{equation}
  \begin{split}
    \left(\frac{dr}{d\xi}\right)^{2}+\mathcal{V}_{eff}(r) & =\frac{c^2}{b^{2}} \\   \mathcal{V}_{eff}(r) & =\frac{N(r)c^{4}}{r^{2}} ,\qquad b^2=\frac{\bar{\mathcal{J}}^2}{\bar{\mathcal{E}}^2}
    \label{y1}
  \end{split}
\end{equation}
considering the chain-rule $\frac{dr}{d\xi}\frac{d\xi}{d\varphi}=\frac{dr}{d\varphi}$ we have the polar equation and the deflection of a light-ray which comes from infinity, pass near to black hole $R_{0}$ and return to infinity
\begin{equation}
    \left(\frac{dr}{d\varphi}\right)^{2}  =\frac{r^{4}}{b^{2}c^{2}}-r^{2}N,\,\,\,\,
    \Delta\varphi =2\int_{R_{0}}^{\infty}
    \frac{1}{r^{2}}\qty(\frac{1}{b^{2}c^{2}}-\frac{N}{r^{2}})^{-1/2}
    \label{desvi1}
\end{equation}
at boundary ($r\rightarrow\infty$)  that polar equation can be solved $r\varphi=bc$, from that, $b$ is interpreted as an impact parameter. 
The right-hand side equation (\ref{desvi1}) 
is very useful in order to get the angle of 
deflection of light rays coming from infinity and passing close to the black hole $r=R_{0}$,
where $R_{0}$ comes from $\eval{\frac{dr}{d\xi}}_{R_{0}}=0$ or $\eval{\frac{dr}{d\varphi}}_{R_{0}}=0$; in the Schwarzschild case that equation is $R_{0}^{3}-b^{2}c^{2}R_{0}-b^{2}c^{2}r_{h}=0$.

In the literature we usually have the following change
of radial coordinate $r(\varphi)=1/u(\varphi)$, in the left-hand side equation of (\ref{desvi1}), in order
to get the second order equation. Finally, we can transform it in two differential equations in order to apply the Runge-Kutta method to solve it.
\begin{figure}
\begin{subfigure}{.5\textwidth}
	\centering
	\includegraphics[width=.8\linewidth]{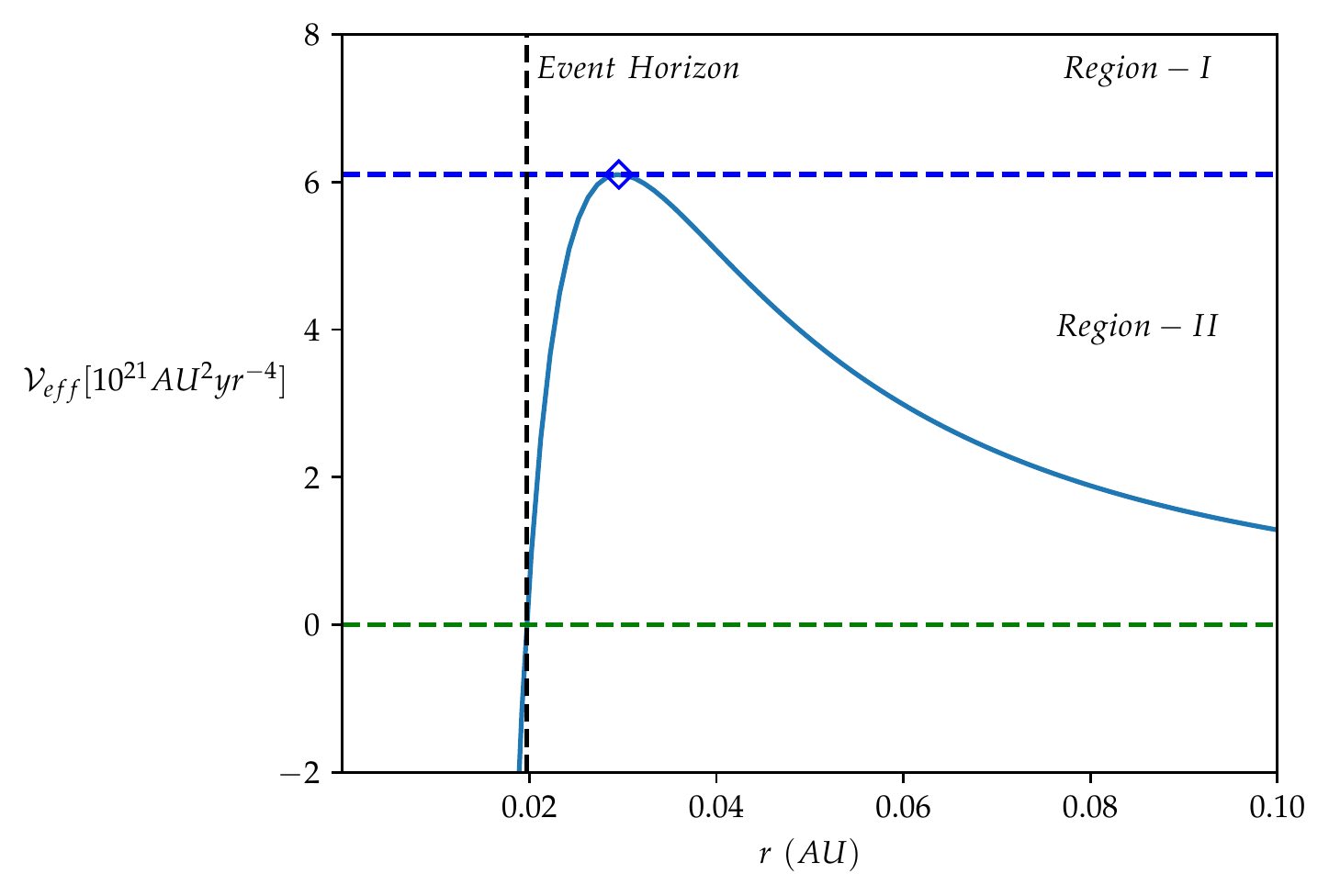}
	\caption{}
	\label{fig17a}
\end{subfigure}%
\begin{subfigure}{.5\textwidth}
	\centering
	\includegraphics[width=.8\linewidth]{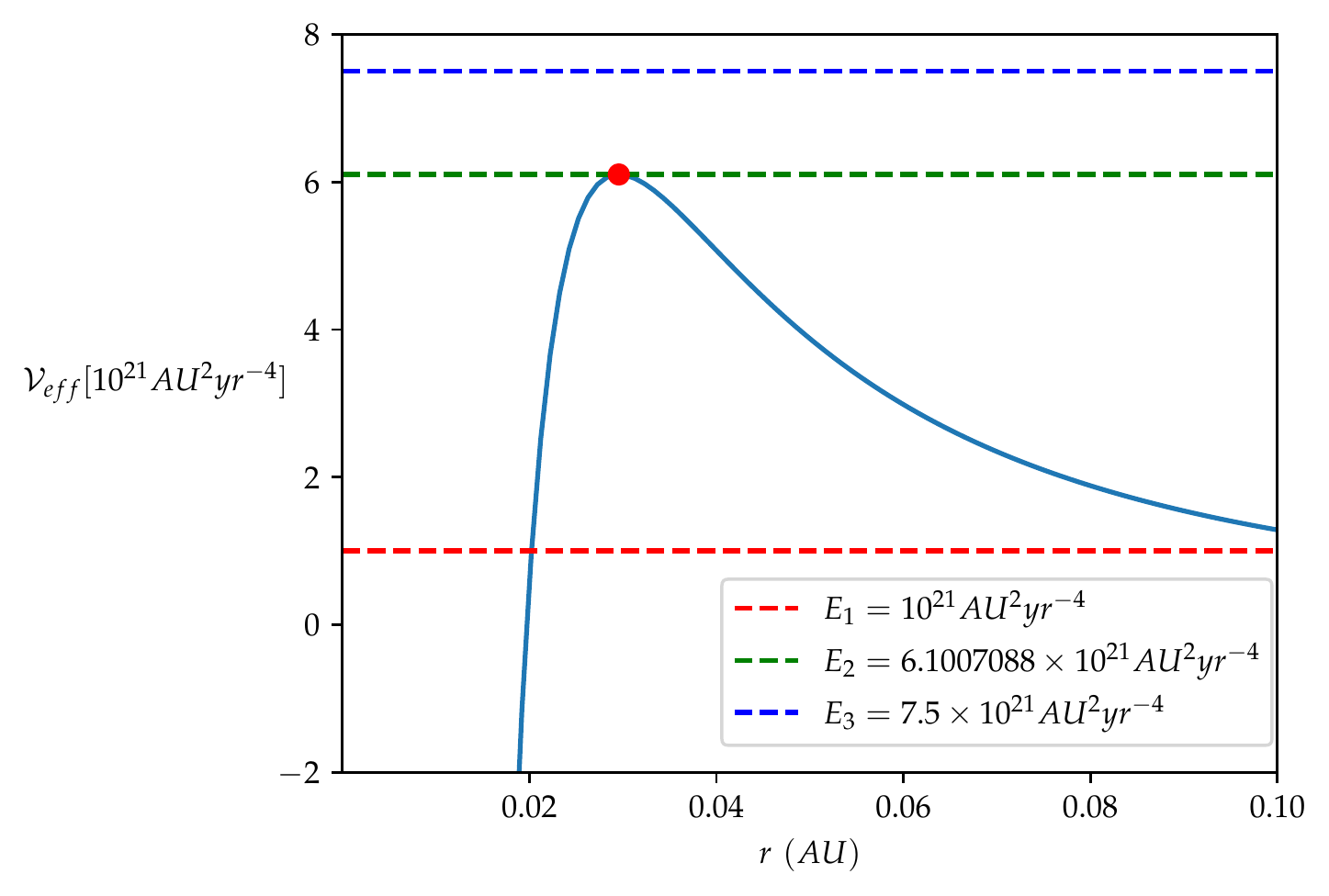}
	\caption{}
	\label{fig17b}
\end{subfigure}
\caption{\small (\subref{fig17a}):Region I, Above $E=\mathcal{V}^{max}_{eff}$. Region II: Between $E=0$ and $E=\mathcal{V}^{max}_{eff}$. To make this plot we have considered $M=10^6 M_\odot$.\\
(\subref{fig17b}): $E_1$ belongs to Region II. $E_2$ corresponds to the maximum of the effective potential. $E_3$ belongs to the region I.}
\label{fig17}
\end{figure}

\begin{figure}
\begin{subfigure}{.5\textwidth}
	\centering
	\includegraphics[width=.8\linewidth]{fig/figE21.pdf}
	\caption{}
	\label{fig18a}
\end{subfigure}%
\begin{subfigure}{.5\textwidth}
	\centering
	\includegraphics[width=.8\linewidth]{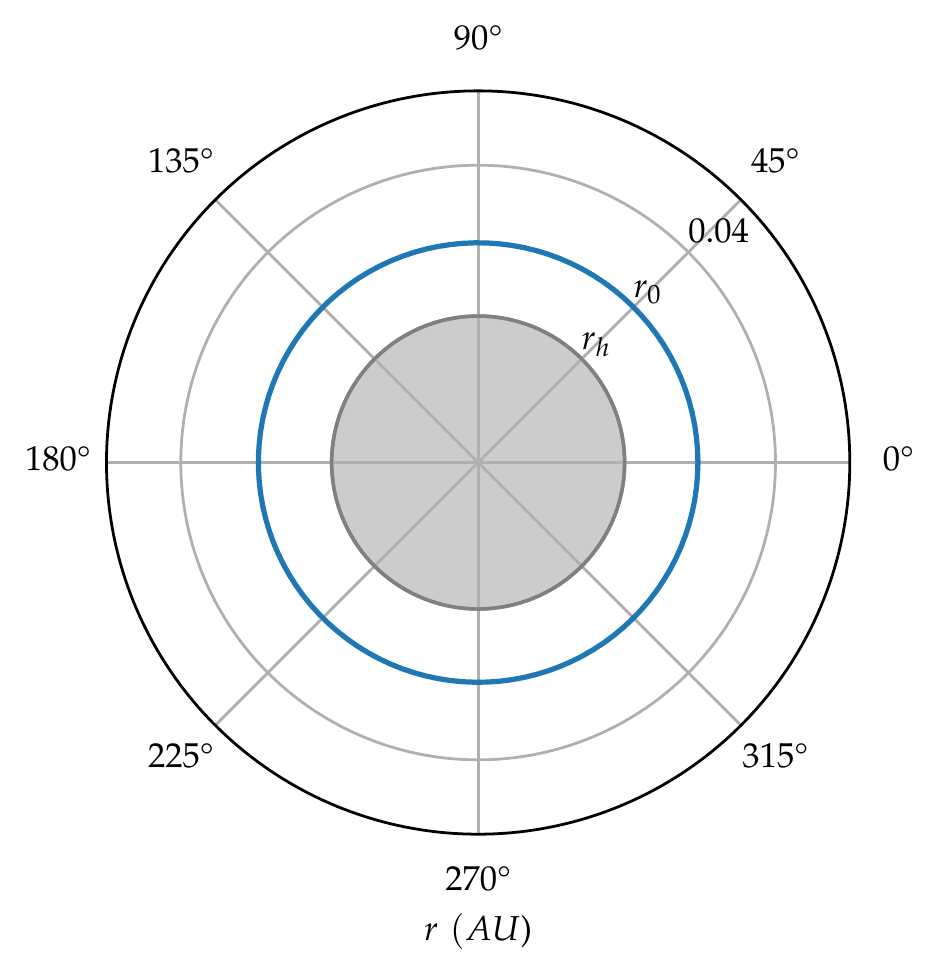}
	\caption{}
	\label{fig18b}
\end{subfigure}
\begin{subfigure}{.5\textwidth}
	\centering
	\includegraphics[width=.8\linewidth]{fig/figE23.pdf}
	\caption{}
	\label{fig18c}
\end{subfigure}
\caption{\small\textbf{ (Null-like)}. \\
(\subref{fig18a}): $E=E_1=10^{21} AU^2 yr^{-4}$, initial conditions: $r_0\rightarrow\infty$ and $\left(\frac{dr}{d\xi}\right)_0=\sqrt{E}$.\\
(\subref{fig18b}): $E=E_2=\mathcal{V}_{eff}^{max}=\mathcal{V}_{eff}(r_*)$, initial conditions: $r_0=r_*$ and $\left(\frac{dr}{d\xi}\right)_0=0$.\\
(\subref{fig18c}):$E=E_1=7.5\times 10^{21} AU^2 yr^{-4}$, initial conditions: $r_0\rightarrow\infty$ and $\left(\frac{dr}{d\xi}\right)_0=\sqrt{E}$.}
\label{fig:fig}
\end{figure}

\setcounter{equation}{0}

%%%%%%%%%%%%
%%%%%%%%%%%%%%%%%%%%%%%%%%%%%%%%%%%%%%%%%%%%%%%%%%%%%%%%%%%%%%
\section{Near-horizon geometry of Schwarzschild black hole}
\label{nearSch}
%%%%%%%%%%%
%
The present section is an interesting example which can help us to understand the hairy case.
The near horizon geometry of Schwarzschild black hole (\ref{Sch1}) is such that
$r=r_{h}+\epsilon$
\begin{equation}
  N(r)=1-\frac{r_{h}}{r}~\Rightarrow~
  N(r_{h}+\epsilon)\approx(r-r_{h})N^{'}(r_{h})
\end{equation}
the metric is given by
\begin{equation}
  ds^{2}=-\frac{(r-r_{h})}{r_{h}}c^{2}dt^{2}+\frac{r_{h}dr^{2}}{(r-r_{h})}+r^{2}_{h}(d\theta^{2}+\sin^{2}{\theta}d\varphi^{2})
  \label{eqq1}
\end{equation}
The Lorentzian signature of the metric imposes the condition $r\geq r_{h}$. 
Choosing a new radial coordinate $\rho^{2}=4r_{h}(r-r_{h})$ we get the Rindler geometry
%
%%%%%%%%%%%%%%%%%%%%%%
\subsection{Time-like geodesics}
%%%%%%%%%%%%%%%%%%%%%%
%
From (\ref{eqq1}) the equation for time-like geodesics near to black hole is 
\begin{equation}
  - c^2
  =-N(r)c^{2}\dot{t}^{2}+\frac{\dot{r}^{2}}{N(r)}+r_{h}^{2}\dot{\varphi}^{2}, \qquad N(r)=\frac{(r-r_{h})}{r_{h}}
\end{equation}
\begin{align}
   & g_{tt}\xi_{(t)}^{t}u^{t}=-\frac{\mathcal{E}}{m}
  \qquad\Rightarrow\qquad 
  \frac{dt}{d\tau}=\left(\frac{\mathcal{E}}{mc^2}\right)
  \frac{1}{N(r)}=\frac{r_{h}\bar{\mathcal{E}}}{r-r_{h}}     \\ 
   & g_{\varphi\varphi}\xi_{(\varphi)}^{\varphi}u^{\varphi}
  =\frac{\mathcal{J}}{m}
  \qquad\Rightarrow\qquad
  \frac{d\varphi}{d\tau}=\left(\frac{\mathcal{J}}{mc^{2}}\right)\frac{c^{2}}{r_{h}^{2}}=\frac{\bar{\mathcal{J}}c^{2}}{r_{h}^{2}}
\end{align}
It describes the radial motion of a test body with energy $E=\bar{\mathcal{E}}^2-1$ in the effective potential
\begin{equation}
  E=\frac{\dot{r}^{2}}{c^{2}}+U_{\text{eff}}(r), \qquad 
  U_{\text{eff}}(r)=\frac{(r-r_{h})}{r_{h}}\left(1+\frac{\bar{\mathcal{J}}^{2}c^{2}}{r_{h}^{2}}\right)-1
  \label{poo}
\end{equation}
The polar equation can be constructed 
considering the chain-rule $\frac{dr}{d\tau}=\frac{dr}{d\varphi}\frac{d\varphi}{d\tau}$, from that we have
\begin{equation}
  \qty(\frac{dr}{d\varphi})^{2}=
  \frac{r_{h}^{4}}{\bar{\mathcal{J}}^{2}c^{2}}(E-U_{\text{eff}})
  \label{devi2}
\end{equation}
The typical second order orbital equation
can be constructed taking the the derivative  (\ref{devi2}) with respect to $\varphi$. The orbital equation and the solution are
\begin{equation}
  \begin{split}
    \dv[2]{r}{\varphi} & =-\frac{(\bar{\mathcal{J}}^{2}c^{2}+r_{h}^{2})r_{h}}{2c^{2}\bar{\mathcal{J}}^{2}} \\
    ~\Rightarrow~
    r(\varphi) & =-\frac{r_{h}}{4}\qty(1+\frac{r_{h}^{2}}{\bar{\mathcal{J}}^{2}c^{2}})\varphi^{2}+c_{1}\varphi+c_{2}
    \label{sol1}
  \end{split}
\end{equation}
The initial condition at the horizon for $(dr/d\varphi)_{r_{h}}$ can be obtained from (\ref{poo}), and at horizon we fix $r(\varphi)$ as
\begin{equation}
  \qty(\dv{r}{\varphi})_{r_{h}}=\frac{r_{h}^{2}}{\bar{\mathcal{J}}c}\sqrt{E+1}=c_{1}, \qquad r(\varphi=0)=r_{h}=c_{2}
\end{equation}
then, the solution (\ref{sol1}) under the above initial conditions is given by
\begin{equation}
  r(\varphi)=-\frac{r_{h}}{4}\qty(1+\frac{r_{h}^{2}}{\bar{\mathcal{J}}^{2}c^{2}})\varphi^{2}+\qty(\frac{r_{h}^{2}}{\bar{\mathcal{J}}c}\sqrt{E+1})\varphi+r_{h}
\end{equation}
%
%%%%%%%%%%%%%%%%%%%%%%
\subsection{Null geodesic}
%%%%%%%%%%%%%%%%%%%%%%
%
We consider an affine parameter $\xi$ and scale it as
$\xi\rightarrow \xi/\bar{\mathcal{J}}$
\begin{equation}
  \frac{dt}{d\xi}=
  \frac{r_{h}\bar{\mathcal{E}}}{\bar{\mathcal{J}}(r-r_{h})}, \qquad
  \frac{d\varphi}{d\xi}=\frac{c^{2}}{r_{h}^{2}}
\end{equation}
in (\ref{eqq1}) at $ds^{2}=0$ we get
\begin{equation}
  \begin{split}
    \left(\frac{dr}{d\xi}\right)^{2}+ & \mathcal{V}_{eff}(r)=\frac{c^2}{b^{2}}, \\   
    \mathcal{V}_{eff}(r) & =\frac{(r-r_{h})c^{4}}{r_{h}^{3}} ,\qquad b^2=\frac{\bar{\mathcal{J}}^2}{\bar{\mathcal{E}}^2}
    \label{y1}
  \end{split}
\end{equation}
considering the chain-rule $\frac{dr}{d\xi}\frac{d\xi}{d\varphi}=\frac{dr}{d\varphi}$ we obtain the polar equation
\begin{equation}
  \left(\frac{dr}{d\varphi}\right)^{2}=\frac{r_{h}^{4}}{b^{2}c^{2}}-r_{h}(r-r_{h})
  \label{desvi3}
\end{equation}
Taking the derivative 
of the orbital equation and integrating we have
\begin{equation}
  \dv[2]{r}{\varphi}=-\frac{r_{h}}{2} ~\Rightarrow~
  r(\varphi)=-\frac{r_{h}\varphi^{2}}{4}+c_{1}\varphi+c_{2}
\end{equation}
The initial conditions 
are: from (\ref{desvi3}) we get $(dr/d\varphi)_{r_{h}}$, and fixing that $r(\varphi=0)=r_{h}$
\begin{equation}
  \qty(\dv{r}{\varphi})_{r_{h}}=\frac{r_{h}^{2}}{bc}=c_{1}, \qquad r(\varphi=0)=r_{h}=c_{2}
\end{equation}
we have
\begin{equation}\label{b10}
  r(\varphi)=-\frac{r_{h}}{2}\varphi^{2}+\frac{r_{h}^{2}}{bc}\varphi+r_{h}
\end{equation}
\begin{figure}
    \centering
    \includegraphics[width=0.7\linewidth]{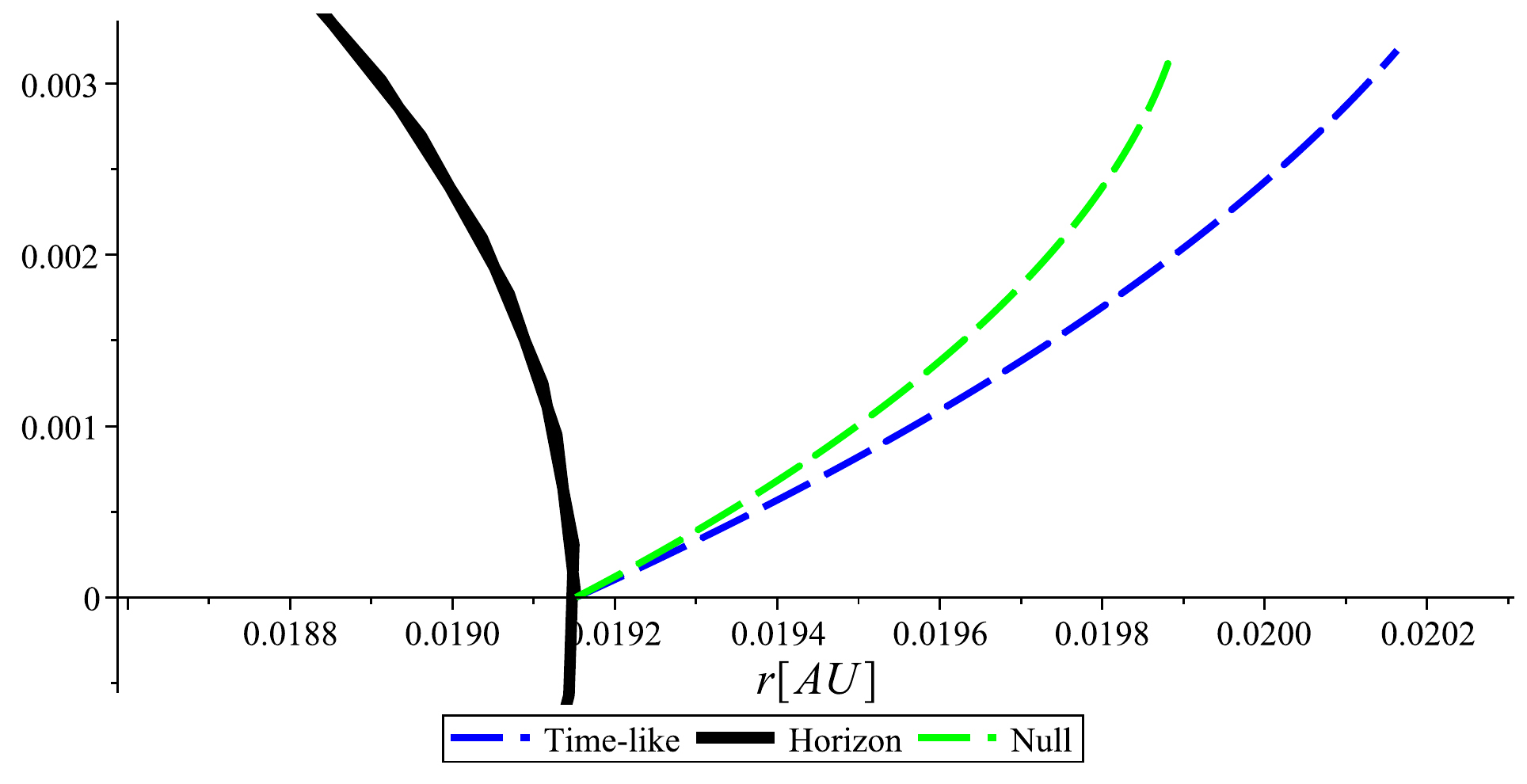}
  \caption{\textbf{Near horizon geodesics}.
  Here we plot the near horizon geodesics of the figure \ref{timee:fig2} and figure \ref{time14:2}, which are null and time-like respectively. Here we consider the angular momentum per unit mas $\bar{\mathcal{J}}=7.322\times 10^{-7}$ and the mass of the black hole $M=10^{6}M_\odot$. For time-like geodesic we consider $E=0.3$ and for null case $b^{2}=5.47329\times 10^{-13} yr$, $E=7.5\times 10^{21} AU^{2}yr^{-4}$. The black line define the horizon of the black hole of radius $r_{h}=0.0192~AU$. The Schwarzschild black hole has a geodesics which fall inside of the black hole with some angle with respect to the tangent to horizon.} 
  \label{SchNear}
\end{figure}

\end{appendices}

% \hypersetup{linkcolor=blue}
% \phantomsection % use it for correct TOC link !!!
% \addtocontents{toc}{\protect\addvspace{4.5pt}}% add vertical space in TOC
% \addcontentsline{toc}{section}{References} 

\newpage
\bibliography{hairyBH}

\end{document}